\documentclass{article}

\usepackage[utf8]{inputenc}
\usepackage[english]{babel}
\usepackage[top=3cm, bottom=3.5cm, right=3cm, left=3cm]{geometry}
\usepackage{hyperref}
\usepackage[x11names]{xcolor}
\usepackage{amssymb}
\usepackage{amsmath}
\usepackage{amsthm}
\usepackage{amsfonts}
\usepackage{mathtools}
\usepackage{mathabx}
\usepackage{esvect}
\usepackage{graphicx}
\usepackage{stmaryrd}
\usepackage{mathrsfs}
\usepackage{tipa}
\usepackage{array}
\usepackage{tabu}
\usepackage{colortbl}
\usepackage{url}
\usepackage{wasysym}
\usepackage{paralist}

\usepackage{tikz}
\usetikzlibrary{positioning} 
\usetikzlibrary{shapes} 
\usetikzlibrary{arrows,arrows.meta,decorations.markings} 
\tikzstyle directed=[postaction={decorate,decoration={markings,mark=at position .65 with {\arrow[scale=1.2]{>}}}}]

\newtheorem{theorem}{Theorem}
\newtheorem{lemma}{Lemma}

\newcommand{\zero}{\textsf{0}}
\newcommand{\one}{\textsf{{1}}}
\newcommand{\Ctwo}{\ensuremath{\mathscr{C}_r}}
\newcommand{\Cone}{\ensuremath{\mathscr{C}_\ell}}
\newcommand{\com}{\ensuremath{c}}
\newcommand{\allzero}{\ensuremath{(\zero^\ell,\zero^r)}}
\newcommand{\allone}{\ensuremath{(\one^\ell,\one^r)}}
\newcommand{\altzero}{\ensuremath{
	((\zero\one)^{\frac{\ell}{2}},(\zero\one)^{\frac{r}{2}})}}
\newcommand{\altone}{\ensuremath{
	((\one\zero)^{\frac{\ell}{2}},(\one\zero)^{\frac{r}{2}})}}
\newcommand{\cop}[2]{\texttt{\small copy}(#1, #2)}
\newcommand{\pcop}{\texttt{\small copy}}
\newcommand{\copc}[3]{\texttt{\small copy\_c}(#1, #2, #3)}
\newcommand{\pcopc}{\texttt{\small copy\_c}}
\newcommand{\copp}[2]{\texttt{\small copy\_p}(#1, #2)}
\newcommand{\pcopp}{\texttt{\small copy\_p}}
\newcommand{\sync}{\texttt{\small sync}}
\renewcommand{\clock}[3]{\texttt{\small incUp}(#1, #2, #3)}
\newcommand{\pclock}{\texttt{\small incUp}}
\newcommand{\counter}[3]{\texttt{\small decUp}(#1, #2, #3)}
\newcommand{\pcounter}{\texttt{\small decUp}}
\newcommand{\shift}[1]{\texttt{\small shift}(#1)}
\newcommand{\pshift}{\texttt{\small shift}}
\newcommand{\update}[1]{\texttt{\small update}(#1)}
\newcommand{\pupdate}{\texttt{\small update}}
\newcommand{\erase}[1]{\texttt{\small erase}(#1)}
\newcommand{\perase}{\texttt{\small erase}}
\newcommand{\expand}[1]{\texttt{\small expand}(#1)}
\newcommand{\pexpand}{\texttt{\small expand}}

\newcommand{\size}[1]{\texttt{\small size}(#1)}

\newcommand{\seq}[1]{\texttt{\small sequence}(#1)}

\newcommand{\convp}[1]{\texttt{\small fix1}(#1)}
\newcommand{\pconvp}{\texttt{\small fix1}}
\newcommand{\convn}[1]{\texttt{\small fix0}(#1)}
\newcommand{\pconvn}{\texttt{\small fix0}}
\newcommand{\simp}[1]{\texttt{\small simp}(#1)}
\newcommand{\psimp}{\texttt{\small simp}}
\newcommand{\compa}[1]{\texttt{\small comp1}(#1)}
\newcommand{\pcompa}{\texttt{\small comp1}}
\newcommand{\compb}[1]{\texttt{\small comp2}(#1)}
\newcommand{\pcompb}{\texttt{\small comp2}}
\newcommand{\comp}[1]{\texttt{\small comp}(#1)}
\newcommand{\pcomp}{\texttt{\small comp}}

\DeclareMathOperator{\euler}{\ensuremath{\phi}}
\DeclareMathOperator{\id}{\ensuremath{\textsf{\footnotesize id}}}

\DeclareMathOperator{\inv}{\ensuremath{\textsf{\footnotesize inv}}}
\DeclareMathOperator{\convol}{\ensuremath{\bigstar}}

\DeclareMathOperator{\perrin}{{\tt P}}
\DeclareMathOperator{\lucas}{{\tt L}}
\DeclareMathOperator{\gold}{\ensuremath{\varrho}}
\DeclareMathOperator{\plastic}{\ensuremath{\xi}}

\def\quant{\ensuremath{\texttt{Q}}}

\def\NBrec{\ensuremath{\texttt{X}}}
\def\NBrecp{\ensuremath{\texttt{X}^+}}
\def\NBrecn{\ensuremath{\texttt{X}^-}}

\def\NBrecnp{\ensuremath{\texttt{X}^{-,+}}}
\def\NBrecnn{\ensuremath{\texttt{X}^{-,-}}}
\def\NBrecmin{\ensuremath{\widetilde{\texttt{X}}}}

\def\NBatt{\ensuremath{\texttt{A}}}
\def\NBattp{\ensuremath{\texttt{A}^+}}
\def\NBattn{\ensuremath{\texttt{A}^-}}

\def\NBtotatt{\ensuremath{\texttt{T}}}
\def\NBtotattp{\ensuremath{\texttt{T}^+}}
\def\NBtotattn{\ensuremath{\texttt{T}^-}}

\newcommand{\NN}{\mathbb{N}}
\newcommand{\ZZ}{\mathbb{Z}}
\newcommand{\BB}{\mathbb{B}}

\newcommand{\ANs}{automata networks}
\newcommand{\BAN}{Boolean automata network}
\newcommand{\BANs}{Boolean automata networks}
\newcommand{\BAC}{Boolean automata cycle}
\newcommand{\BACs}{Boolean automata cycles}
\newcommand{\BADC}{Boolean automata double-cycle}
\newcommand{\BADCs}{Boolean automata double-cycles}

\def\Un{\ensuremath{\mbox{1\hspace{-.2em}l}}}
\def\un{\ensuremath{\textsf{\footnotesize one}}}
\renewcommand{\bar}[1]{\overline{#1}}

\begin{document}

\title{Boolean automata isolated cycles and tangential double-cycles dynamics}

\author{
  Jacques Demongeot~$^{1,}$\thanks{
		\href{mailto:jacques.demongeot@univ-grenoble-alpes.fr}{jacques.demongeot@univ-grenoble-alpes.fr}
	}{~}, 
  Tarek Melliti~$^{2,}$\thanks{
		\href{mailto:tarek.melliti@univ-evry.fr}{tarek.melliti@univ-evry.fr}
	}{~},
  Mathilde Noual~$^{3,}$\thanks{
		\href{mailto:mathilde.noual@incaseofpeace.com}{mathilde.noual@incaseofpeace.com}
	}{~},\\
  Damien Regnault~$^{2,}$\thanks{
		\href{mailto:damien.regnault@univ-evry.fr}{damien.regnault@univ-evry.fr}
	}{~} and
  Sylvain Sen{\'e}~$^{4,}$\thanks{
		\href{mailto:sylvain.sene@univ-amu.fr}{sylvain.sene@univ-amu.fr}
	}\\[2mm]
	{\small $^1$~Université publique, Grenoble, France}\\
	{\small $^2$~Université publique, Évry, France}\\
	{\small $^3$~Freie Universität Berlin, Germany}\\
	{\small $^4$~Université publique, Marseille, France}\\
}

\date{}

\maketitle

\begin{abstract}{Abstract}
	Our daily social and political life is more and more impacted by social networks. 
  The functioning of our living bodies is deeply dependent on biological regulation networks such as neural, genetic, and protein networks. 
  And the physical world in which we evolve, is also structured by systems of interacting particles. 
  Interaction networks can be seen in  all spheres of existence that concern us,  and yet, our understanding of interaction networks remains severely limited by our present lack of both theoretical and applied insight into their clockworks.
  In the past, efforts at understanding interaction networks have mostly been directed towards applications. 
  This has happened at the expense of developing understanding of the generic and fundamental aspects of interaction networks (properties and behaviours due primarily to the fact that a system is an interaction network, as opposed to properties and behaviours rather due to the fact a system is a \textit{genetic} interaction network for instance).
  Intrinsic properties of interaction networks (\emph{e.g.}, the ways in which they transmit information along  entities, their ability to produce this or that kind of global dynamical behaviour depending on local interactions) are thus still not well understood. 
  Lack of fundamental knowledge tends to limit the innovating power of applications. 
  Without more theoretical fundamental knowledge, applications cannot evolve deeply and become more impacting.
  Hence, it is necessary to better apprehend and comprehend the intrinsic properties of interaction networks, notably the relations between their architecture and their dynamics and how they are affected by and set in \emph{time}.
  In this chapter, we use the elementary mathematical model of \BANs{} as a formal archetype of interaction networks. 
  We survey results concerning the role of feedback cycles and the role of  intersections between feedback cycles, in shaping the asymptotic dynamical behaviours of interaction networks. 
  We pay special attention to the impact of the automata updating modes.\\[2mm]
\end{abstract}

\section{Introduction}
\label{sec:intro}

Interaction networks occupy an important place in our daily life. 
We see this today in particular with the massive use of social networks, the fundamental implications and mechanisms of which we hardly have any understanding of.
And social media is just one example among many other kinds of interaction networks that affect us consequentially.
At all levels of our lives there are interaction networks, that can be comprised as sets of entities that interact locally with each other over time. 

In an interaction network, local interactions take place. And as a result of these, the network as a whole exhibits  global behaviours that generally remain  difficult to explain on the sole basis of local processes. 
Undeniably one of the most telling examples lies at the origin of all living organisms: genes and their regulation, associated with other mechanisms inducing variability (splicing, role of the chromatin, \emph{etc.}). 
It is currently accepted in biology and medicine, that a better understanding of genetic regulation is a necessary condition for improving our knowledge of life, which in turn could allow us to achieve a more precise comprehension of pathologic mechanisms, and to access more targeted and thereby more efficient therapies.\smallskip

Whilst the application aspects of interaction networks are obviously of real and tangible importance, and they have progressed quantitatively at a frantic pace over the last twenty years, the more fundamental aspects aiming at understanding and analysing the intrinsic properties of these networks have so far received less attention from the  scientific community. 
Current applications, in particular those emerging in biology (the privileged domain of application of the present chapter), are awaiting significant theoretical advances to continue their qualitative development. 
Undoubtedly, such theoretical advances may be the fruit of the combination of computer science and discrete mathematics.
Together with bioinformatics, viewed here as ``the study of computer processes in biotic systems''~\cite{J-Hesper1970,J-Hogeweg1978}, they seem particularly suited to meet the needs arising from applications.
Indeed, real networks, through the entities and interactions that compose them, can naturally be viewed as discrete objects. 
They can easily be represented by computer models which capture their 
essence thanks to a high level of abstraction. 
Where traditional (continuous) mathematical modelling focuses mainly on the quantitative characteristics of networks, the interest of discrete computer science modelling comes from the qualitative nature of the questions that it raises, which makes it possible to realise that the central elements of networks are not the entities themselves but rather the interactions that link them.\smallskip

This survey adopts the qualitative point of view of fundamental computer science to examine some properties of interaction networks. 
This choice of approach is consistent with the origins of modern computing. 
Indeed, investigations of formal neural networks~\cite{J-McCulloch1943} and cellular automata~\cite{L-vonNeumann1966} in the 1940s, both strongly inspired by natural processes, helped establish the first links between data processing and biology. 
Also, the pioneering work of McCulloch and Pitts introduced automata networks as a fundamental model of interaction networks. 
For reasons discussed later, we choose to rely on the same model here.\smallskip

We thus here study automata networks, and more specifically \BANs{}, from a fundamental point of view interested in developing a qualitative understanding of networks. 
We focus on understanding how feedback cycles that are part of network architectures, come to influence the asymptotic dynamical properties of these networks.
By way of methods from discrete dynamical systems theory, enumerative combinatorics, algorithms and complexity theory, we explore the  behavioural diversity of cycles and their tangential intersections.\smallskip

Another point of focus in this chapter is how updating modes influence cycles and cycle intersections (and this network behaviours). 
From both the fundamental and the applied points of view, updating modes are known to be of decisive importance in the shaping of a network's behaviour.
Updating modes define the way in which the states of the automata of a network are updated as a function of a discrete time. 
The time being unbounded towards the future, there is an infinite number of different possible updating modes which we can come with and update the automata of a network according to.
We focus here on the two modes that are the most customary in the literature: the parallel mode and the asynchronous mode. 
The parallel mode is a deterministic and periodic mode. At each time step, it updates the states of all the automata in a network.
The asynchronous mode is a non-deterministic mode. It allows  for any possible series of sequential updates -- which update exactly one automaton per time step -- to take place.
We show how the differences between these two updating modes imply profound differences in the network dynamics.\smallskip

This chapter is a synthesis of results about interaction cycles and variations of interaction cycles derived since the early 2000s. 
The list of results presented here is not exhaustive. 
Details and demonstrations of these results can be found in the literature~\cite{J-Demongeot2012,C-Melliti2015,T-Noual2012,J-Remy2003,T-Sene2012}.\medskip

We will start with a presentation of the history of automata networks in the context of modern fundamental computer science, including a mention of the links that this science has always entertained with biology. 
Main definitions and notations will be explained after that. 
Then, we  will briefly present seminal results obtained in the 1980s that  highlight the crucial importance of the role of feedback cycles on the dynamical expressiveness and behavioural diversity of interaction networks. 
This will lead us to focus on the dynamical and combinatorial properties of isolated cycles when they are subjected to the parallel and asynchronous updating modes. 
And finally, before concluding, we will take a step towards contextualising cycles and look into tangential intersections of cycles, called ``tangential double-cycles'' for short.

\section{Automata networks: between fundamental computer science and biology}
\label{sec:ANs}

Automata network research is part of natural computation field in computer science. 
And the scope of natural computation is twofold. 
First, it designs and develops models of computations that draw inspiration from natural phenomenology. 
Second, it manipulates such models so as to build up a firm, albeit necessarily incomplete, grasp of biological reality natural phenomena. 
As mentioned above, automata networks were initially introduced as a theoretical model of neural networks. 
And since the late 1960s, the literature evidences that they are also relevant as models of genetic regulation networks.
In this section, we provide a comprehensive overview of the interconnections between computer science and biology manifested through \ANs{}.

\subsection{Biology as an inspiration for modern computer science}
\label{subsec:ANs:bio}

Generally, when we are interested in the history of the so-called modern computer science, we go back to the 1930s. 
This period saw the development of classical computing paradigms such as the recursive functions of Herbrand and G{\"o}del, the main works of which are available in~\cite{CL-Godel1934}, the lambda-calculus developed initially in~\cite{J-Church1932}, and also the Turing machines~\cite{J-Turing1936}. 
However, to think of modern computer science only in terms of these paradigms is to forget a whole family of less conventional models that have  grounded many developments in computer science of relevance today.
This is the family of automata networks that we mainly owe to McCulloch and Pitts, and von Neumann, whose first elements of the theory date from the 1940s. 
Based on advances at the time, the original works~\cite{J-McCulloch1943,L-vonNeumann1966} highlight a desire to develop the science of computation while inspiring and advancing the modelling of natural biological phenomena. 
Thus, McCulloch and Pitts introduced the model of formal neural networks which provides an abstraction of neural interactions. 
They showed in particular that propositional logic can represent neural events and that these networks can be considered, to a certain extent, as a universal model of 
computation. 
A little later, in the late 1940s, von Neumann developed cellular automata in order to ``compare natural and artificial automata'' and ``abstract the logical structure of life''. 
The result of his work was the construction of the first self-reproducing and universal cellular automaton.\smallskip

Formal neural networks and cellular automata constitute the founding base of the theory of automata networks, an automata network being defined ``roughly'' as a set of entities (automata) which interact with each other. Their interactions happen in a discrete time, according to  transition functions which are local to the entities. The two major differences that distinguish these two models relate to the number of interacting entities and the nature of the network on which the entities are placed. 
Indeed,  cellular automata are defined by default as having an \textit{infinite} number of entities (a.k.a.  cells). And the entities are placed on a \textit{regular and homogeneous} network (in general $\mathbb{Z}^d$, with $d > 0$). Formal neural networks (also called threshold \BANs{}) have a \textit{finite} number of entities. And the entities are placed on an \textit{irregular and heterogeneous} network. These differences give to each of these models its own characteristics that have enabled strong advances in computer science.\smallskip

By their infinite nature, cellular automata have been mainly studied for their computability properties. 
Among them, we find the Turing universality of the self-reproducing automaton~\cite{L-vonNeumann1966}, of the Game of Life of Conway~\cite{L-Conway1982} and of the elementary cellular automaton 110~\cite{J-Cook2004}. 
Also, in 1971, Smith showed that any Turing machine could be simulated by a cellular automaton defined on $\mathbb{Z}$~\cite{J-Smith1971}. 
In addition, cellular automata proved to be good mathematical tools for studying the parallel functioning of computers, of which they were at the origin of systolic architectures~\cite{CL-Kung1980}. 
Finally, within the framework of dynamical system theory, the desire to understand  their behavioural diversity was at the origin of studies of complexity~\cite{J-Wolfram1984,J-Kurka1997}. 
Formal neural networks have also brought a lot of progress. 
In their original article, McCulloch and Pitts showed that they can simulate any Boolean function. 
Kleene resumed this work. Based on their finite nature, he proved that the languages recognised by these objects are regular, which also allowed him to introduce the concept of finite automata~\cite{CL-Kleene1956}. 
Behavioural characterisation conditions were also given and algebraic methods were then developed within this framework~\cite{J-Huffman1959,J-Elspas1959,L-Golomb1967,J-Cull1971}. 
This last reference highlights in particular, strong links between these networks and the Boolean model of genetic regulation networks introduced in~\cite{J-Kauffman1969b,J-Kauffman1969}. 
This latter model  is at the origin of numerous works emphasising the interest taken by computer science in the context of research in theoretical biology.

\subsection{Computer science as a methodological source for biology}
\label{subsec:ANs:cs}

Understanding the mechanisms of biological regulation, in all their diversity, is one of the major current problems in molecular biology. This was notably highlighted by Jacob and Monod in the early 1960s, in particular in~\cite{J-Jacob1961,J-Monod1963}. 
However, from the end of the 1960s, an observation was shared by two biologists, Kauffman (biochemist and biophysicist) and Thomas (biochemist and geneticist). 
The usual methods of treatment stemming from molecular biology are not adapted to treat as a whole such a problem at the genetic level. 
According to them, the experimental nature of the methods specific to biology at the time can only provide a piecemeal response to this problem and needs to be supplemented by methodological approaches. 
On the basis of Delbr{\"u}ck's work, according to which there are links between differentiated cell types and the attractors of theoretical network models~\cite{C-Delbruck1949}, Kauffman and Thomas proposed to use discrete mathematics to go beyond simple observational knowledge of regulatory systems, advocating in a sense that biology must move towards more general and systematic approaches to living things. 
This resulted in two articles which organise and federate a whole section of research at the frontier between discrete mathematics and theoretical biology~\cite{J-Kauffman1969b,J-Thomas1973}.\smallskip

Kauffman is thus the first to have proposed a model of genetic regulatory 
networks, based on formal neural networks. 
This model is known as \emph{Boolean networks}. It is a formalisation of regulation where genes are the vertices of a randomly constructed graph. Genes interact over time (discrete). Their interactions are dictated by local Boolean transition functions. They determine whether the genes can be expressed or not, that is to say, transcribed or not~\cite{CL-Kauffman1971}. 
Originally, this model is based on two strong hypotheses: the interactions are based on an architecture of $k$-regular graphs, namely graphs of which all the vertices have the same number of neighbours; the evolution is perfectly synchronous (or parallel).  
Relaxations of these hypotheses were subsequently carried out~\cite{J-Aldana2003,C-Gershenson2004}. They  gave rise to applications to biological problems such as the analysis of the behaviour of the yeast regulatory network~\cite{J-Kauffman2003}, and more generally to the analysis of signalling networks~\cite{J-Gupta2007,J-Albert2009}. 
In 1973, Thomas opposed the parallelism hypothesis and developed another method that sought to be closer to ``genetic reality''~\cite{J-Thomas1973}. 
This method comes with two new ideas. The first proposes to represent the causal dynamics of genetic regulations by means of an \textit{asynchronous} state transition system. The second idea is to represent the networks themselves by digraphs whose arcs are signed according to the promoting or inhibiting nature of interactions.
In~\cite{J-Thomas1981}, Thomas introduces two fundamental conjectures, proven in the discrete framework in~\cite{J-Richard2007,J-Richard2010}. 
The first one (resp. The second one) states that the presence of a positive cycle (resp. of a negative cycle), composed of an even number (resp. an odd number) of inhibitory edges, in the architecture of the network is necessary for dynamical multi-stationarity (resp. for the existence of an oscillating limit regime). 
Beyond theoretical research, this method has been widely applied in biology, such as for example the immune response~\cite{J-Kaufman1985,J-Mendoza2006,J-Saez2007} or to infection of \emph{Escherichia coli} by $\lambda$ phage ~\cite{J-Thomas1995,J-Thieffry1995,J-Guet2002}.\smallskip

Although they were initiated by scientists from biology, these two visions emphasise the relevance of \ANs{} and thus bear the mark of computer science and discrete mathematics. 
The contribution of mathematician  F. Robert played a key role. 
From the end of the 1960s he pioneered the study of automata networks from a more fundamental and formal point of view. 
Indeed,  Kauffman and Thomas made ``arbitrary'' choices regarding the ways of 
updating the automata over time. But Robert was interested in the updating modes as 
such and their influences on the networks behaviours. 
He formalised the concepts of block-sequential iterations and chaotic iterations~\cite{J-Robert1969,J-Robert1976,J-Robert1980,L-Robert1986,L-Robert1995}, which  make it possible to obtain updating modes that are partly synchronous and asynchronous. 
This makes perfect sense in theoretical biology since there is no biological argument today to define the temporal organisation of genetic regulations. 
In addition, the work carried out by Robert and his collaborators made it possible to establish solid theoretical bases (simple and general) for the behavioural study of automata networks~\cite{J-Goles1982,J-Cosnard1985,J-Goles1985,L-Goles1990} while keeping in mind their representational capacities for biology~\cite{T-Demongeot1975,L-Cosnard1983,L-Demongeot1985}. 
In this theoretical framework, Robert proved the essential role played by cycles in defining the intrinsic behavioural properties of networks: his theorem stipulates that any acyclic network has a trivial behaviour and admits to the temporal asymptote only a single point fixed. 
Again, Robert's work has found many applications in biology, including modelling the genetic control of the flower development of \emph{Arabidopsis thaliana}~\cite{J-Mendoza1998,J-Demongeot2010,C-Ruz2018} and the study of ventral invagination during gastrointestinal morphogenesis in Drosophila~\cite{J-Aracena2006}.

\subsection{\BANs{}, a simple but complex model}
\label{subsec:ANs:BANs}

From a general point of view, \ANs{} can be used to model any system which satisfies the following three properties: 
\begin{itemize}
\item It is made up of distinct entities that interact with each other; 
\item Each entity is characterised by a variable quantity, which precisely calls 
  to be translated in terms of states of the corresponding automaton in the model; 
\item The events undergone by the system, like the mechanisms that are at their 
  origin, cannot be observed directly or integrally with certainty. Only their 
  consequences are, that is, changes that are fully accomplished. 
\end{itemize}
These three properties impose very few restrictions on the set of systems that can be abstracted and thus modelled by \ANs{}. 
These theoretical objects are therefore generic models of a very wide variety of real systems. 
It is therefore quite easy to understand the reasons that pushed scientists to use them and to keep studying them in the context of ``fundamental bioinformatics''.\smallskip

Let us return to entities' characteristic "variable quantity" mentioned above. 
Translating the quantity  in terms of automata states, calls for a  first exercise of formalisation. 
This consists in choosing whether what interests us in the variation of the 
quantity is of a Boolean, discrete or continuous nature. 
As an illustration, let us take the example of genetic regulation and choose the action of a gene as a variable quantity. 
If, in the action of this gene, what interests us is its expression (and its non-expression), then the state of the automaton chosen to model the gene should  be Boolean. 
If it is the different ways that this gene has of acting on the other elements of the system that interests us, then we can choose to match an automaton state with each way.
This induces a discrete formalism which can obviously be encoded without loss in a Boolean formalism, since an automaton with $k$ states can be represented by $\log_2 (k)$ Boolean automata. 
Finally, if we measure the action of the gene by means of the concentration of 
proteins it produces, continuous formalism turns out to be the most natural. 
On the other hand, it brings a quantitative character. 
If this aspect is not desired, the tendency will then be to approximate the protein concentration function at intervals in order to fall back into a discrete, even Boolean framework by considering only extreme concentrations for example. 
We can therefore grant different statuses to the Boolean context depending on whether we see it as a direct modelling of reality or as an approximation or encoding of an intrinsically continuous or discrete modelling. 
Note that direct Boolean modelling is consistent with the choice to focus on the state changes of the automata rather than on their states themselves. 
By analogy to mechanics, if we see automata as internal combustion engines, the interest relates to the fact that an engine is capable of going from the ``off'' state to the ``on'' state (and vice versa) rather than the amount of electricity supplied by the battery to start or on that released by the candles to cause the explosion and initiate the movement. 
Under this assumption, Boolean abstraction is necessary and sufficient. Furthermore, in order to place ourselves in the context of modelling in biology, it should be emphasised that the discourse of biologists is generally imbued with syntactic elements of propositional logic. 
It is not uncommon to hear sentences such as: ``in the absence of the repressor $\alpha$, the $\beta$ gene is expressed'' or even ``if the products of the $\alpha$ and $\beta$ genes form a complex, the latter promotes the expression of the $\gamma$ gene while these genes tend to inhibit its expression when they are in monomeric form''. 
This syntax also fits perfectly with a direct modelling of reality in the Boolean formalism.\smallskip

In addition, \BANs{} derive other interesting benefits from their simplicity of definition. 
In particular, they provide a framework with clearly defined contours, ideal for tackling fundamental problems around the modelling of interacting entity systems. 
Given the variety of their nature and the current state of our knowledge, the problems in question could not currently benefit from significantly more elaborate frameworks. 
This would inevitably lead to delaying the initial questions and to destructuring the problem posed by paying attention to additional problems induced by the set of parameters to be considered and not intrinsically included in the initial problem. 
For these issues, on the contrary, \BANs{} offer only what is essential and facilitate the manipulation of a minimal concept of causality, which is rooted in the notion of state changes. 
Their merit therefore lies in the reliability of the information they potentially provide, delivered by their very high level of abstraction that also makes it possible to obtain general laws that remain valid in more specific contexts. 
In other words, it is crucial to understand that this simplicity of definition does not necessarily make them ``simplistic'', and does not detract from their ability to model complex phenomena that they allow to analyse qualitatively with a surprising subtlety.

\section{General definitions and notations}
\label{sec:def_nota}

Informally, a \BAN{} is comprised of abstract entities that interact with each other. The abstract entities are called automata. Automata have states which  can take one of two values: either $\zero$ (inactive) or $\one$ (active). The states of automata  can change over the course of a discrete time. They change under  the influence of the states of other automata in the network. 
This section aims to present the formalism of this model, giving the main definitions and useful notations in the rest of the chapter.

\subsection{\BANs{}}
\label{sec:def_nota:BANs}

Let $\BB = \{\zero, \one\}$ and $V = \{0, \dots, n-1\}$ be a set of $n$ Boolean 
automata such that $\forall i \in V$, $x_i\in \BB$ represents the \emph{state} of 
automaton $i$. 
A \emph{configuration} $x$ of a \BAN{} $f$ of size $n$ assigns a value of $\BB$ to each of the automata of $V$ and is classically noted as a vector $x = (x_0, \dots, x_{n-1})$ that is a vertex of the $n$-cube $\BB^n$, or as a binary word $x = x_0\dots x_{n-1}$. 
Formally, a \emph{\BAN{}} $f$ of size $n$ whose set of automata is $V$ is an ordered set of $n$ Boolean functions, such that $f = (f_i: \BB^n \to \BB\ |\ i \in V)$. 
Given $i \in V$, $f_i$ is the \emph{local transition function} of automaton 
$i$. 
It predetermines its evolution from any configuration $x$: if $i$ is updated in configuration $x$ at time $t$, it goes from state $x_i(t)$ to state $f_i(x(t)) = x_i(t + 1)$.\smallskip

\begin{figure}[t!]
  \begin{center}
    \begin{minipage}{.4\textwidth}
      \centerline{\scalebox{1}{\begin{tikzpicture}[>=latex,auto]
        \tikzstyle{node} = [circle, thick, draw]
        \node[node](n0) at (0,0) {$0$};
        \node[node](n1) at (0,2) {$1$};
        \node[node](n2) at (2,1) {$2$};
        \draw[thick, ->] (n0) edge [loop left] node {$+$} (n0);
        \draw[thick, ->] (n0) edge node [swap] {$-$} (n1);
        \draw[thick, ->] (n0) edge [bend right] node [swap] {$-$} (n2);
        \draw[thick, ->] (n1) edge [bend right]  node [swap] {$-$} (n0);
        \draw[thick, ->] (n1) edge [loop left]  node {$-$} (n1);
        \draw[thick, ->] (n1) edge node [swap] {$-$} (n2);
        \draw[thick, ->] (n2) edge [bend right] node [swap] {$+$} (n1);
        \draw[thick, ->] (n2) edge [loop right] node {$-$} (n2);
      \end{tikzpicture}}}
    \end{minipage}
    \qquad
    \begin{minipage}{.4\textwidth}
      \centerline{$\left\lbrace \begin{array}{l}
        f_0(x) = x_0 \lor \neg x_1\\
        f_1(x) = \neg x_0 \lor \neg x_1 \lor x_2\\
        f_2(x) = \neg x_0 \lor \neg x_1 \lor \neg x_2\\
      \end{array}\right.$}
    \end{minipage}
    \caption{A \BAN{} of size $3$: its interaction graph (on the left), 
       the ordered set of its local transition functions (on the right).}
    \label{fig:ban}
  \end{center}
\end{figure}
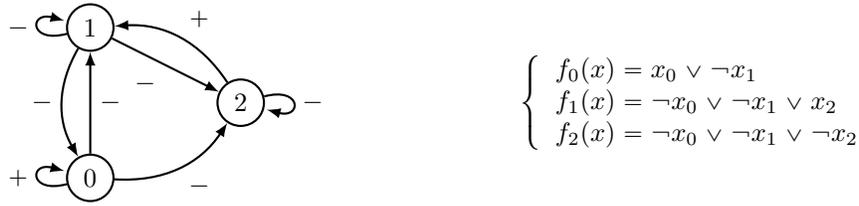

Let $s: \BB \to \Un$, with $\Un = \{-1, 1\}$, defined such that $s(b) = b - (\neg b)$, the function allowing to convert a Boolean number into a signed integer in $\Un$. 
In this chapter, we pay particular attention to the state changes of automata, which leads us to introduce the following notations for all $x$ in $\BB^n$:
\begin{equation*}
	\forall i \in V,\ \bar{x}^{\{i\}} = (x_0, \dots, x_{i-1}, \neg x_{i}, x_{i+1}, 
	\dots, x_{n-1})\text{.}
\end{equation*}
\begin{equation*}
	\forall W \subseteq V,\ \forall i \notin W,\ \bar{x}^{W \cup \{i\}} = 
	\bar{\bar{x}^W}^{\{i\}}\text{.}
\end{equation*}
The \emph{sign of an interaction} from $i$ to $j$ in configuration $x$ is defined 
by $\text{sign}_x(i, j) = s(x_i) \cdot (f_j(x) - f_j(\bar{x}^{\{i\}}))$. 
The \emph{effective interactions} in $x$ belong to $E(x) = \{(i, j) \in V \times V\ |\ \text{sign}_x(i, j) \neq 0\}$. 
From there, we define the \emph{interaction graph} of $f$ as being the oriented graph $G = (V, E)$, where $E = \bigcup_{x \in \BB^n} E(x)$ is the set of interactions. 
In this chapter, the \BANs{} discussed (see section 3.3) are special in the sense that their interaction graphs are simple, namely that there can only be one signed interaction $(i, j) \in E$. 
If it is signed positively (resp. negatively), we say that it is activating (resp. inhibiting) and the state of $j$ tends to mimic (resp. to oppose) that of $i$. 
In the following, the interaction graphs will be signed for convenience of reading (see Figure~\ref{fig:ban}).

\subsection{Updating modes and transition graphs}
\label{sec:def_nota:UM_TG}

In order to determine the possible behaviours of a \BAN{}, it is essential to specify the way according to which the states of the automata (or, abusing language, the automata) are updated over time. 
This specification is what we call an updating mode. 
The most general point of view is to consider all the possibilities. 
This amounts to seeing the evolution of a network as a discrete dynamical system associated with a relation so that, for each configuration, $2^{n} - 1$ outgoing transitions are taken into account, namely a transition for each subset of automata whose states can be updated. 
More precisely, for all $W \neq \emptyset \subseteq V$, we define the update function $F_W: \BB^n \to \BB^n$ such that:
\begin{equation*}
	\forall x \in \BB^n, \forall i \in V,\ F_W(x)_i = \begin{cases}
		f_i(x) & \text{if } i \in W\text{,}\\
		x_i & \text{otherwise.}
	\end{cases}
\end{equation*}
Thus, for the most general updating mode, called the \emph{elementary updating mode}, the global behaviour of the network is given by the elementary transition graph $\mathscr{G}_e = (\BB^n, T_e)$, where $T_e = \{(x, F_W(x))\ |\ x \in \BB^n, W \neq \emptyset \subseteq V\}$, introduced in~\cite{T-Noual2012,T-Sene2012}.\smallskip

\begin{figure}[t!]
  \begin{center}
    \begin{minipage}{.45\textwidth}
      \centerline{\scalebox{1}{\begin{tikzpicture}[>=to,auto]
        \tikzstyle{conf} = [rectangle, draw]
        \tikzstyle{pf} = [rectangle, thick, draw, fill=black!20]
        \tikzstyle{lc} = [rectangle, thick, draw, fill=black!80]
        \node[conf](n000) at (0,0) {$\zero\zero\zero$};
        \node[conf](n001) at (2.5,0) {$\zero\zero\one$};
        \node[conf](n010) at (0,2.5) {$\zero\one\zero$};
        \node[pf](n011) at (2.5,2.5) {$\zero\one\one$};
        \node[lc](n100) at (1.25,1.25) {\textcolor{white}{$\one\zero\zero$}};
        \node[lc](n101) at (3.75,1.25) {\textcolor{white}{$\one\zero\one$}};
        \node[lc](n110) at (1.25,3.75) {\textcolor{white}{$\one\one\zero$}};
        \node[lc](n111) at (3.75,3.75) {\textcolor{white}{$\one\one\one$}};
        \draw[thick,Green4,->] (n000) edge (n001);
        \draw[thick,Blue3,->] (n000) edge (n010);
        \draw[thick,Red3,->] (n000) edge (n100);
        \draw[thick,Green4,->] (n001) edge [loop right, distance=5mm] (n001);
        \draw[thick,Blue3,->] (n001) edge (n011);
        \draw[thick,Red3,->] (n001) edge (n101);
        \draw[thick,Green4,->] (n010) edge (n011);
        \draw[thick,Blue3,->] (n010) edge [loop left, distance=5mm] (n010);
        \draw[thick,Red3,->] (n010) edge [loop left, distance=7mm] (n010);
        \draw[thick,Green4,->] (n011) edge [loop right, distance=5mm] (n011);
        \draw[thick,Blue3,->] (n011) edge [loop right, distance=7mm] (n011);
        \draw[thick,Red3,->] (n011) edge [loop right, distance=9mm] (n011);
        \draw[thick,Green4,->] (n100) edge (n101);
        \draw[thick,Blue3,->] (n100) edge (n110);
        \draw[thick,Red3,->] (n100) edge [loop left, distance=5mm] (n100);
        \draw[thick,Green4,->] (n101) edge [loop right, distance=5mm] (n101);
        \draw[thick,Blue3,->] (n101) edge (n111);
        \draw[thick,Red3,->] (n101) edge [loop right, distance=7mm] (n101);
        \draw[thick,Green4,->] (n110) edge (n111);
        \draw[thick,Blue3,->] (n110) edge (n100);
        \draw[thick,Red3,->] (n110) edge [loop left, distance=5mm] (n110);
        \draw[thick,Green4,->] (n111) edge (n110);
        \draw[thick,Blue3,->] (n111) edge [loop right, distance=5mm] (n111);
        \draw[thick,Red3,->] (n111) edge [loop right, distance=7mm] (n111);
      \end{tikzpicture}}}
    \end{minipage}
    \quad\vrule\quad
    \begin{minipage}{.45\textwidth}
      \centerline{\scalebox{1}{\begin{tikzpicture}[>=to,auto]
        \tikzstyle{conf} = [rectangle, thick, draw]
        \tikzstyle{pf} = [rectangle, thick, draw, fill=black!20]
        \tikzstyle{lc} = [rectangle, thick, draw, fill=black!80]
        \node[conf](n000) at (0,0) {$\zero\zero\zero$};
        \node[conf](n001) at (2.5,0) {$\zero\zero\one$};
        \node[conf](n010) at (0,2.5) {$\zero\one\zero$};
        \node[pf](n011) at (2.5,2.5) {$\zero\one\one$};
        \node[conf](n100) at (1.25,1.25) {$\one\zero\zero$};
        \node[lc](n101) at (3.75,1.25) {\textcolor{white}{$\one\zero\one$}};
        \node[lc](n110) at (1.25,3.75) {\textcolor{white}{$\one\one\zero$}};
        \node[lc](n111) at (3.75,3.75) {\textcolor{white}{$\one\one\one$}};
        \draw[thick,->] (n000) edge [bend left] (n111);
        \draw[thick,->] (n001) edge (n111);
        \draw[thick,->] (n010) edge (n011);
        \draw[thick,->] (n011) edge [loop right, distance=5mm](n011);
        \draw[thick,->] (n100) edge [bend left] (n111);
        \draw[thick,->] (n101) edge (n111);
        \draw[thick,->] (n110) edge [bend right] (n101);
        \draw[thick,->] (n111) edge (n110);
      \end{tikzpicture}}}
    \end{minipage}
    \caption{Two transition graphs of the \BAN{} defined in Figure~\ref{fig:ban}. 
      Left panel: its asynchronous transition graph where every $\textcolor{Red3}{\to}$ (resp. $\textcolor{Blue3}{\to}$ and $\textcolor{Green4}{\to}$) represents an update of automaton $0$ (resp. $1$ and $2$).  
      Right panel: its parallel transition graph. 
      In each graph, stable configurations (a.k.a. fixed points) are depicted in light gray while recurring configurations belonging to stable oscillations (a.k.a. limit cycles) are depicted in dark gray.}
    \label{fig:tg}
  \end{center}
\end{figure}
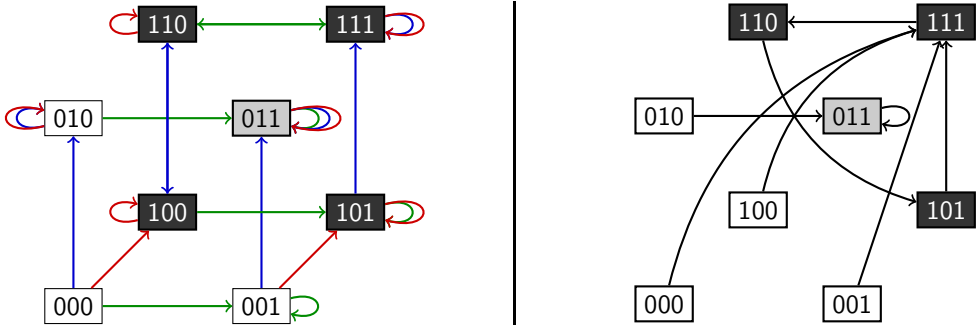

The transitions $(x, F_{i}(x))$ that involve updating a single automaton $i \in V$ are called \emph{asynchronous transitions}. 
The transitions $(x, F_W(x))$, with $|W| > 1$, that induce the updating of several automata are called \emph{synchronous transitions}. 
The subgraph $\mathscr{G}_a = (\BB^n, T_a)$ of $\mathscr{G}_e$ whose set of arcs $T_a = \{(x, F_{\{i\}}(x))\ |\ x \in \BB^n, i \in V\}$ equals the set of asynchronous transitions of network is called the \emph{asynchronous transition graph}. 
This graph defines the asynchronous dynamics of the network that corresponds to its dynamics when it evolves according to the asynchronous updating mode, \emph{i.e.} such that in each configuration, only $n$ transitions are considered, one for each automaton. 
This updating mode has been widely used in the studies of Thomas and his collaborators~\cite{J-Remy2003,J-Remy2008,J-Richard2004,J-Richard2007,J-Richard2010,J-Thomas1981,J-Thomas1991}. 
As an illustration, the asynchronous transition graph of the network depicted in Figure~\ref{fig:ban} is presented in Figure~\ref{fig:tg} (left). 
However, because it is the most ``natural'' (mathematically speaking) when only the local transition functions are known, and because that it allows to give relevant insights through transition graphs of smaller sizes, the \emph{parallel updating mode} occupies a very important place in the literature on discrete dynamical systems in general. 
When a network evolves in parallel, all of its automata update their states at 
each time step. 
In other words, the parallel transition graph of a network is $\mathscr{G}_p = (\BB^n, T_p)$, where $T_p = \{(x, F_V(x)), x \in \BB^n\}$ is the subset of the perfectly synchronous transitions of $T_e$. 
The parallel transition graph of the network given in Figure~\ref{fig:ban} is presented in Figure~\ref{fig:tg} (right).

Let us now specify notations and vocabulary relative to  dynamical behaviours of networks. 
Consider an arbitrary \BAN{} $f$ of size $n$, an updating mode $\mu$, the associated transition graph $\mathscr{G}_\mu = (\BB^n, T_\mu)$, and $x \in \BB^n$ one of its possible configurations. 
A \emph{trajectory} of $x$ is any path in $\mathscr{G}_\mu$ that starts from $x$. 
A strongly connected component of $\mathscr{G}_\mu$ which does not admit any outgoing transition is an asymptotic behaviour of $(f, \mu)$, that we classically designate as an \emph{attractor} of $(f, \mu)$. 
A configuration of $\BB^n$ that belongs to an attractor is a \emph{recurring 
configuration}. 
Given an attractor, its \emph{length} is the number of recurring configurations that compose it. 
An attractor of length $1$ (resp. of length strictly greater than $1$) is a \emph{stable configuration}, a.k.a. a \emph{fixed point} (resp. a \emph{stable oscillation}, a.k.a. a \emph{limit cycle}) of $(f, \mu)$. 
If $\mu$ is a deterministic updating mode, such as the parallel mode, attractors are simple cycles and the term \emph{period} is preferred to refer to their length. Finally, we define the \emph{convergence time of a configuration} $x$ as the length of its shortest trajectory which makes it reach a recurring configuration. 
The  \emph{convergence time of the network} is the greatest convergence time of all 
its $2^n$ configurations. 
In the transition graphs presented in this chapter, by convention, we associate the 
light gray color with stable configurations and the dark gray color with recurrent 
configurations belonging to stable oscillations. 
Thus, in Figure~\ref{fig:tg}, it can be seen that configuration $\zero\one\one$ is 
stable for the asynchronous and parallel updating modes. 
This is a direct consequence of the fact that a stable configuration that is a fixed point of $f$ (implicitly evolving in parallel) is preserved by every updating mode since it corresponds to the vector of the local fixed points of the local transition functions. 
On the other hand, we observe that the asynchronous transition graph admits a stable oscillation of size $4$ while the parallel update mode admits a stable oscillation of size $3$. 
This illustrates that stable oscillations are generally not conserved when the updating mode is changed. This is a current important field of study in the context of interaction networks and \BANs{}.

\subsection{Isolated cycles and tangential cycles}
\label{sec:def_nota:IC_TC}

As mentioned in the introduction, we focus in this chapter on two sorts of  \BANs{}, namely, \textit{\BACs{}}  and    \textit{\BADCs{}}. The first  are networks whose interaction graphs are cycles. The second are networks whose interaction graphs are two cycles that intersect tangentially. 
The founding results that lead us to develop ever more research on these interaction patterns are presented in the following section.\smallskip

A \emph{\BAC{}} $\mathscr{C}_n$ is a \BAN{} of size $n$ whose interaction graph 
$G = (V, E)$ is a cycle, in the sense of graph theory. 
$V$ is naturally assimilated to $\ZZ / n\ZZ$, so that considering two automata $i$ and $j$ of $V$, $i + j$ represents $i + j \mod n$. 
Thus, a \BAC{} $\mathscr{C}_n$ is defined as an ordered set of local transition functions of arity $1$ that are such that: $\forall i \in V,\ f_i: \BB^n \to \BB$, and either $f_i(x) = x_{i-1}$ or $f_i(x) = \neg x_{i-1}$. 
Note that there are two types of \BAC{}, positive and negative. 
A \BAC{} is a \emph{positive cycle} $\mathscr{C}^+$ (resp. a \emph{negative cycle} $\mathscr{C}^-$) if it is composed of an even number (resp. of an odd number) of inhibiting interactions.\smallskip

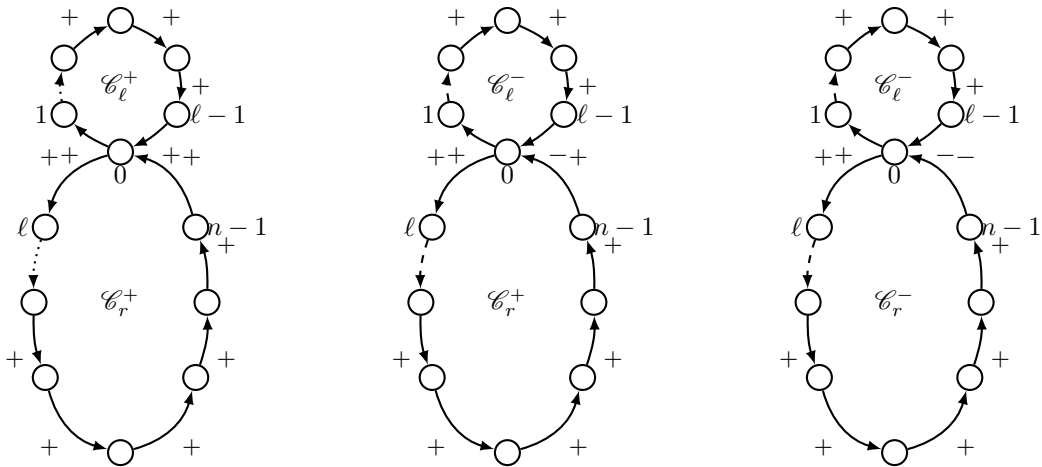
\begin{figure}[t!]
  \begin{center}
    \begin{minipage}{.3\textwidth}
      \centerline{\scalebox{1}{\begin{tikzpicture}[>=latex,auto]
        \tikzstyle{node} = [circle, thick, draw]
        \tikzstyle{mknode} = []
        \node[mknode] (ml) at (0,0.875) {$\mathscr{C}_\ell^+$};
        \node[mknode] (mr) at (0,-2) {$\mathscr{C}_r^+$};
        \node[node] (0) at (0,0) {};
        \node[mknode] (m0) at (0,-.3) {$0$};
        \node[node] (1) at (-.75,.5) {};
        \node[mknode] (m1) at (-1.05,.5) {$1$};
        \node[node] (2) at (-.75,1.25) {};
        \node[node] (3) at (0,1.75) {};
        \node[node] (4) at (.75,1.25) {};
        \node[node] (5) at (.75,.5) {};
        \node[mknode] (m5) at (1.3,.5) {$\ell-1$};
        \node[node] (6) at (-1,-1) {};
        \node[mknode] (m5) at (-1.3,-1) {$\ell$};
        \node[node] (7) at (-1.15,-2) {};
        \node[node] (8) at (-1,-3) {};
        \node[node] (9) at (0,-4) {};
        \node[node] (10) at (1,-3) {};
        \node[node] (11) at (1.15,-2) {};
        \node[node] (12) at (1,-1) {};
        \node[mknode] (m12) at (1.55,-1) {$n-1$};
        \draw[thick,->] (0) edge [bend left=10] node {$+$} (1);
        \draw[dotted,thick,->] (1) edge [bend left=10] (2);
        \draw[thick,->] (2) edge [bend left=10] node {$+$} (3);
        \draw[thick,->] (3) edge [bend left=10] node {$+$} (4);
        \draw[thick,->] (4) edge [bend left=10] node {$+$} (5);
        \draw[thick,->] (5) edge [bend left=10] node {$+$} (0);
        \draw[thick,->] (0) edge [bend right] node [swap] {$+$} (6);
        \draw[dotted,thick,->] (6) edge [bend right=10] (7);
        \draw[thick,->] (7) edge [bend right=10] node [swap] {$+$} (8);
        \draw[thick,->] (8) edge [bend right] node [swap] {$+$} (9);
        \draw[thick,->] (9) edge [bend right] node [swap] {$+$} (10);
        \draw[thick,->] (10) edge [bend right=10] node [swap] {$+$} (11);
        \draw[thick,->] (11) edge [bend right=10] node [swap] {$+$} (12);
        \draw[thick,->] (12) edge [bend right] node [swap] {$+$} (0);
      \end{tikzpicture}}}
    \end{minipage}
    \quad
    \begin{minipage}{.3\textwidth}
      \centerline{\scalebox{1}{\begin{tikzpicture}[>=latex,auto]
        \tikzstyle{node} = [circle, thick, draw]
        \tikzstyle{mknode} = []
        \node[mknode] (ml) at (0,0.875) {$\mathscr{C}_\ell^-$};
        \node[mknode] (mr) at (0,-2) {$\mathscr{C}_r^+$};
        \node[node] (0) at (0,0) {};
        \node[mknode] (m0) at (0,-.3) {$0$};
        \node[node] (1) at (-.75,.5) {};
        \node[mknode] (m1) at (-1.05,.5) {$1$};
        \node[node] (2) at (-.75,1.25) {};
        \node[node] (3) at (0,1.75) {};
        \node[node] (4) at (.75,1.25) {};
        \node[node] (5) at (.75,.5) {};
        \node[mknode] (m5) at (1.3,.5) {$\ell-1$};
        \node[node] (6) at (-1,-1) {};
        \node[mknode] (m5) at (-1.3,-1) {$\ell$};
        \node[node] (7) at (-1.15,-2) {};
        \node[node] (8) at (-1,-3) {};
        \node[node] (9) at (0,-4) {};
        \node[node] (10) at (1,-3) {};
        \node[node] (11) at (1.15,-2) {};
        \node[node] (12) at (1,-1) {};
        \node[mknode] (m12) at (1.55,-1) {$n-1$};
        \draw[thick,->] (0) edge [bend left=10] node {$+$} (1);
        \draw[dashed,thick,->] (1) edge [bend left=10] (2);
        \draw[thick,->] (2) edge [bend left=10] node {$+$} (3);
        \draw[thick,->] (3) edge [bend left=10] node {$+$} (4);
        \draw[thick,->] (4) edge [bend left=10] node {$+$} (5);
        \draw[thick,->] (5) edge [bend left=10] node {$-$} (0);
        \draw[thick,->] (0) edge [bend right] node [swap] {$+$} (6);
        \draw[dashed,thick,->] (6) edge [bend right=10] (7);
        \draw[thick,->] (7) edge [bend right=10] node [swap] {$+$} (8);
        \draw[thick,->] (8) edge [bend right] node [swap] {$+$} (9);
        \draw[thick,->] (9) edge [bend right] node [swap] {$+$} (10);
        \draw[thick,->] (10) edge [bend right=10] node [swap] {$+$} (11);
        \draw[thick,->] (11) edge [bend right=10] node [swap] {$+$} (12);
        \draw[thick,->] (12) edge [bend right] node [swap] {$+$} (0);
      \end{tikzpicture}}}
    \end{minipage}
    \quad
    \begin{minipage}{.3\textwidth}
      \centerline{\scalebox{1}{\begin{tikzpicture}[>=latex,auto]
        \tikzstyle{node} = [circle, thick, draw]
        \tikzstyle{mknode} = []
        \node[mknode] (ml) at (0,0.875) {$\mathscr{C}_\ell^-$};
        \node[mknode] (mr) at (0,-2) {$\mathscr{C}_r^-$};
        \node[node] (0) at (0,0) {};
        \node[mknode] (m0) at (0,-.3) {$0$};
        \node[node] (1) at (-.75,.5) {};
        \node[mknode] (m1) at (-1.05,.5) {$1$};
        \node[node] (2) at (-.75,1.25) {};
        \node[node] (3) at (0,1.75) {};
        \node[node] (4) at (.75,1.25) {};
        \node[node] (5) at (.75,.5) {};
        \node[mknode] (m5) at (1.3,.5) {$\ell-1$};
        \node[node] (6) at (-1,-1) {};
        \node[mknode] (m5) at (-1.3,-1) {$\ell$};
        \node[node] (7) at (-1.15,-2) {};
        \node[node] (8) at (-1,-3) {};
        \node[node] (9) at (0,-4) {};
        \node[node] (10) at (1,-3) {};
        \node[node] (11) at (1.15,-2) {};
        \node[node] (12) at (1,-1) {};
        \node[mknode] (m12) at (1.55,-1) {$n-1$};
        \draw[thick,->] (0) edge [bend left=10] node {$+$} (1);
        \draw[dashed,thick,->] (1) edge [bend left=10] (2);
        \draw[thick,->] (2) edge [bend left=10] node {$+$} (3);
        \draw[thick,->] (3) edge [bend left=10] node {$+$} (4);
        \draw[thick,->] (4) edge [bend left=10] node {$+$} (5);
        \draw[thick,->] (5) edge [bend left=10] node {$-$} (0);
        \draw[thick,->] (0) edge [bend right] node [swap] {$+$} (6);
        \draw[dashed,thick,->] (6) edge [bend right=10] (7);
        \draw[thick,->] (7) edge [bend right=10] node [swap] {$+$} (8);
        \draw[thick,->] (8) edge [bend right] node [swap] {$+$} (9);
        \draw[thick,->] (9) edge [bend right] node [swap] {$+$} (10);
        \draw[thick,->] (10) edge [bend right=10] node [swap] {$+$} (11);
        \draw[thick,->] (11) edge [bend right=10] node [swap] {$+$} (12);
        \draw[thick,->] (12) edge [bend right] node [swap] {$-$} (0);
      \end{tikzpicture}}}
    \end{minipage}
    \caption{The three canonical \BADCs{} of size $n = \ell+r-1$: the positive one (on the left),  the mixed one (on the center), and the negative one (on the right).}
    \label{fig:cycles}
  \end{center}
\end{figure}

A \emph{\BADC{}} $\mathscr{D}_n$, with $n = \ell + r - 1$, is a \BAN{} of size $n$ composed of two \BACs{} $\mathscr{C}_\ell$ and $\mathscr{C}_r$ which tangentially intersect in one automaton, automaton $0$. 
In the following, for reasons of clarity, we prefer the notation $\mathscr{D}_{\ell,r}$. 
The set of automata of cycle $\mathscr{C}_\ell$ is $V^\mathcal{L} = \ZZ / \ell\ZZ = \{0, ..., \ell-1\}$ and that of cycle $\mathscr{C}_r$ is $V^\mathcal{R} = {0} \cup \{\ell-1+i\ |\ i \neq 0 \in \ZZ / r\ZZ\}$. 
In a \BADC{}, by definition, all the local transition functions are of arity $1$ except that of automaton $0$ that is of arity $2$. 
In this work, we only consider locally monotonous \BADC{}, which induces that function $f_0$ is defined as $f_0(x) = f_0^\mathcal{L}(x) \diamond f_0^\mathcal{R}(x) = x_{\ell-1} \diamond x_{n-1}$, where $\diamond \in \{\land, \lor\}$. 
Note that the choice of operator $\diamond$ only changes the position of the configurations on the trajectories. 
In other words, whatever the chosen updating mode, given two \BADC{} $\mathscr{D}_{\ell,r}$ and $\mathscr{D}'_{\ell,r}$ such that $f_0(x) = f_{\ell-1}(x) \land f_{n-1}(x)$ and $f'_0(x) = f'_{\ell-1}(x) \lor f'_{n-1}(x)$, their respective transition graphs are identical up to an isomorphism on the configurations. 
In addition, it is trivial to determine one from the other, replacing the configurations with their opposites. 
Consequently, in the rest of this chapter, to insist on this property, we will use without loss of generality $f_0(x) = f_{l-1}(x) \lor f_{n-1}(x)$ (resp. $f_0(x) = f_{\ell-1}(x) \land f_{n-1}(x)$) for \BADCs{} evolving according to the parallel (resp. asynchronous) updating mode. 
Since they are made up of two \BACs{}, it is easy to see that there are three distinct types of \BADCs{}. 
A positive \BADC{} is composed of two positive \BACs{} and is denoted by $\mathscr{D}_{\ell,r}^{+,+}$; a negative \BADC{} is composed of two negative \BACs{} and is denoted by $\mathscr{D}_{\ell,r}^{-,-}$; a mixed \BADC{} is composed of a negative \BAC{} $\mathscr{C}^-$ tangentially intersected with a positive \BAC{} $\mathscr{C}^+$, and is denoted by $\mathscr{D}_{\ell,r}^{-,+}$.\smallskip

Finally, in~\cite{T-Noual2012,T-Sene2012}, the authors have shown that the \BACs{} admit canonical representatives, and \BADCs{} too by induction. 
The studies presented in the sequel focus on these canonical representatives only. 
Indeed, canonicity means that two distinct \BACs{} or \BADCs{} of same sign and same size that evolve following the same updating mode admit the same transition graph up to an isomorphism on the configurations. 
A positive \BAC{} is said to be canonical when its interaction graph contains only activating interactions. 
A negative \BAC{} is canonical when its interaction graph admits a single inhibiting interaction, represented by the arc $(n-1,0) \in E$. 
The canonical \BADCs{} are the canonical \BAC{} compositions, depicted in Figure~\ref{fig:cycles}.

\section{Seminal results on cycles}
\label{sec:semres_cycles}

The works of Robert~\cite{J-Robert1980} and Thomas~\cite{J-Thomas1981} have highlighted three fundamental results that explain the primordial role that cycles play in the behavioural diversity of interaction networks.

\begin{theorem}[\cite{J-Robert1980,L-Robert1986,L-Robert1995}] Let $f: \BB^n \to \BB^n$ be a \BAN{} of size $n$ and $G$ its associated interaction graph. 
  If $G$ is an acyclic graph, then: 
  \begin{enumerate}
  \item $f$ has a unique attractor that is a stable configuration, let us say $x$.
  \item $\mathscr{G}_p$ has a path of length at most $n$ from every configuration $y$ to $x$.
  \item $\mathscr{G}_a$ is acyclic and has a a geodesic path from every configuration $y$ to $x$.
  \end{enumerate}
  \label{th:Robert}
\end{theorem}

This theorem is particularly interesting for two reasons. 
First, it is easy to extend it to any kind of multi-valued automata networks $g: \prod_{i=1}^n X_i \to \prod_{i=1}^n X_i$, where $X_i$ denotes the set of possible states of automaton $i$, and to any kind of updating mode such that every automaton is updated an infinite number of times over the course of time (let us call such an updating mode a fair updating mode). 
Indeed, the general idea of the proof rests on an induction on the depths of the automata of the acyclic interaction graph that admits \textit{source} automata. Source automata are automata that are governed by constant local transition functions.  They  inevitably become forever  fixed once they are updated for the first time. In an acyclic  network, the fixity of source automata propagates. The states of all the other automata progressively become fixed too as a result. 
The theorem emphasizes that cycles are necessary conditions for interaction networks to admit complex dynamics.

\begin{theorem}[\cite{J-Remy2008,J-Richard2007,J-Thomas1981}] Let $g: \prod_{i=1}^n X_i \to \prod_{i=1}^n X_i$ be an automata network and $G$ its associated interaction graph. 
  Under the asynchronous updating mode, the presence of a positive cycle in $G$ is necessary for the dynamics of $g$ to admit several stable configurations.
  \label{th:thomas_cp}
\end{theorem}

This second theorem sheds light on the role of positive cycles on the ability of 
interaction networks to stabilise in several ways. 
Although it was originally stated and demonstrated under the asynchronous updating mode hypothesis, this result has been shown to hold for any fair updating 
mode~\cite{T-Noual2012,T-Sene2012}. 
The general proof starts from the result of~\cite{J-Richard2007} and is made by contradiction under the assumption of the absence of a positive cycle. 
In this case, either the interaction graph is acyclic and Theorem~\ref{th:Robert} applies, or it has at least one negative cycle and it is shown that such cycles, whatever the updating mode, cannot remove the local instabilities on all automata.

\begin{theorem}[\cite{J-Richard2010,J-Thomas1981}] Let $g: \prod_{i=1}^n X_i \to \prod_{i=1}^n X_i$ be an automata network and $G$ its associated interaction graph. 
  Under the asynchronous updating mode, the presence of a negative cycle is necessary for the dynamics of $g$ to admit a stable oscillation.
  \label{th:thomas_cn}
\end{theorem}

As for this third theorem, it turns out not to be general for all updating modes. 
To be convinced, it suffices to compute the parallel transition graph of an arbitrary positive \BAC{} of size greater than $2$ and to note that it admits at least one stable oscillation. 
Despite this lack of generality, it should be noted that this theorem underlines the singular role of negative cycles in connection with asymptotic dynamic oscillations, as will be explained below.\smallskip

Taken together, these three theorems, of which we can say that two of them are laws insofar as their generality makes them sorts of meta-theorems of interaction network theory, emphasise that the feedback cycles are the causes of the dynamical complexity of networks. 
In other words, they effectively constitute the sources of the behavioural diversity of networks and consequently of the computational expressiveness of networks. This naturally brings us  to the following parts of this chapter, about  major results concerning the dynamical and combinatorial properties of \BACs{} and \BADCs{}. 
The results are  illustrated with examples.
As discussed above, Robert's result highlighted feedback cycles as kinds of complexity engines. Then, Thomas' results put forward the necessity to distinguish feedback cycles depending on their (positive or negative) nature to understand their influence. The results presented below about  \BADCs{} are a  first step further, towards  understanding how cycle combinations operate and what their effects are.

\section{\BAC{} dynamics}
\label{sec:BACs}

In this section, we focus on \BACs{}. 
In a first part, we present in Theorem~\ref{th:cycles_p} the main results related to the dynamics of isolated \BACs{} when the latter evolve according to the parallel updating mode. 
This theorem requires some preliminary definitions and notations of number theory such as the Dirichlet convolution, the M{\"o}bius function and the Euler's totient 
function. 
In a second part, we present the results related to isolated \BACs{} when the latter evolve according to the asynchronous updating mode.

\subsection{Parallel \BACs{}}
\label{sec:BACs:p}

The first elements on the dynamics of \BACs{} evolving according to the parallel updating mode were introduced in~\cite{J-Remy2003}. The full characterisation of their dynamics had to wait  until~\cite{J-Demongeot2012}. 
This characterisation could be obtained thanks to an approach combining discrete dynamical systems theory, enumerative and word combinatorics, particularly appropriate to the nature of the mathematical objects in question.

\subsubsection{Definitions and notations of pertinent quantities}

To describe the results obtained, we give below definitions and notations. 
First, given an attractor of (minimal) period $p$, we says that all the multiples of $p$ are also periods of this attractor. 
So if $x \in \BB^n$ is a recurring configuration of an attractor of period $p$, then $p$ is the period of $x$ and of any other configuration $y$ such that there exists $t \in \NN$ such that $y = F_V^t(x)$. 
We denote by $\mathscr{X}(p) = \{x \in \BB^n\ |\ x = F_V^p(x)\}$ the set of recurring configurations of period $p$ and by $X(p) = |\mathscr{X}(p)|$ their number. 
We define the smallest integer $\omega$ that is a period common to all recurring configurations as the \emph{order} of the \BAC{}, which is said \emph{to be reached} when there exists an attractor of minimum period $\omega$.\smallskip

Let us consider the function $\un: n \in \NN \mapsto 1$ as well as the \emph{Dirichlet convolution}, denoted by $\convol$~\cite{L-Apostol1976}. 
Given two functions $f$ and $g$, $\convol$ is the binary operator defined such that $f \convol g: n \in \NN^\ast \mapsto \sum_{p|n} f(p) \cdot g(n / p)$. 
The set of the arithmetic functions with point-to-point addition and Dirichlet convolution is a commutative ring. 
The identity by the multiplication of this ring is the function $\delta:\NN^\ast \to \NN^\ast$, defined by $\delta(1) = 1$ and for all $n > 1$, $\delta(n) = 0$. 
The inverse of the function $\un$ for the Dirichlet convolution is the M{\"o}bius function $\mu$, defined as:
\begin{equation*}
  \mu: n \in \NN^\ast \mapsto \begin{cases}
    0  & \text{if } n \text{ is not square-free,}\\
    1  & \text{if } n > 0 \text{ has an even number of prime factors,}\\
    -1 & \text{if } n > 0 \text{ has an odd number of prime factors.} 
  \end{cases}
\end{equation*}
If $n = \prod_{i = 0}^k p_i$, where the $p_i$s are the distinct prime numbers taken in increasing order, then $\mu(n) = (-1)^k$. 
In our context, this function is of interest through the M{\"o}bius inversion formula that is obtained from $\un \convol \mu = \delta$, that is satisfied by all the functions $f$ and $g$, and that is such that: $g = f \convol \un \implies f = g \convol \mu$. 
In other words, we have:
\begin{equation*}
  \forall n \in \mathbb{N}^\ast,\ g(n) = \sum_{p|n} f(p)\ \implies\ f(n) = \sum_{p|n} g(p) \cdot \mu(n/p)\text{.}
\end{equation*}
Another particularly useful function in the sequel is the \emph{Euler's totient function}, denoted by $\euler$. 
Given an integer $n \in \NN^\ast$, it associates the number of strictly positive integers less than or equal to $n$ that are prime with $n$, such as: $\euler(n) = |\{m \in \NN^\ast\ |\ m \leq n \text{ and } m \text{ is prime with } n\}|$. 
Note that there is a relationship between the M{\"o}bius function and the Euler's totient function. 
Indeed, as $\euler$ satisfies $\forall n \in \NN^\ast,\ n = \euler \convol \textsf{\footnotesize un}(n)$, it respects $\euler = \mu \convol \id$, where $\id: n \in \NN^\ast \mapsto n$.\smallskip

In terms of combinatorics, the asymptotic behaviour of an interaction network can be described by means of four quantities~\cite{J-Puri2001} related to each other  and given below. 
Consider that $p$ is a divisor of the order $\omega$ of the \BAC{} studied and the function $\inv: n \in \mathbb{N}^\ast \to 1/n$. 
We then have the following quantities:
\begin{itemize}
\item The \emph{number} $\NBrec(p)$ \emph{of configurations of period $p$} is $\NBrec = \NBrecmin \convol \un$;
\item The \emph{number} $\NBrecmin(p)$ \emph{of configurations of minimal period $p$} is $\NBrecmin = \NBrec \convol \mu$;
\item The \emph{number} $\NBatt(p) = \NBrecmin(p)/p$ \emph{of attractors of period $p$} is $\NBatt = \inv (\NBrec \convol \mu)$;
\item The \emph{total number} $\NBtotatt(\omega)$ \emph{of attractors} is $\NBtotatt = \NBatt \convol \un = \inv (\NBrec \convol \euler)$.
\end{itemize}
The last two quantities correspond to well known formulas in the context of Lyndon words and binary necklaces~\cite{J-Berstel2007,L-Graham1989,M-Ruskey2003}: a Lyndon word $w$ being such that $w < v$ for all nonempty words $v$ such that $w = uv$ and $u$ is nonempty; a binary necklace $w$ of length $n$ being a circular binary word such that for all $i$ in $\ZZ$, $w_i = w_i \mod n$. 
In particular, the penultimate defining $\NBatt$ corresponds to the Witt formula counting the number of Lyndon words; the last one defining $\NBtotatt$ corresponds to the Burnside's orbit-counting lemma. 
Note that the last formula is satisfied because $\inv$ distributes over $\convol$. 
Finally, note that it suffices to calculate $\NBrec$ to obtain the others.

\subsubsection{Results}

The qualitative characterisation of the dynamics of \BACs{} can be summarised by the following theorem that presents in the form of a table the set of all the quantities presented above. 
For reasons of space, the details and the demonstrations related to these results are not presented here. 
The reader can nevertheless find all the details in~\cite{J-Demongeot2012,T-Noual2012,T-Sene2012}.

\begin{figure}[t!]
  \begin{center}
    \begin{minipage}{.45\textwidth}
      \centerline{\scalebox{1}{\begin{tikzpicture}[>=to,auto]
        \tikzstyle{conf} = [rectangle, thick, draw]
        \tikzstyle{pf} = [rectangle, thick, draw, fill=black!20]
        \tikzstyle{lc} = [rectangle, thick, draw, fill=black!80]
        \node[pf](n000) at (0,0) {$\zero\zero\zero$};
        \node[lc](n001) at (2.5,0) {\textcolor{white}{$\zero\zero\one$}};
        \node[lc](n010) at (0,2.5) {\textcolor{white}{$\zero\one\zero$}};
        \node[lc](n011) at (2.5,2.5) {\textcolor{white}{$\zero\one\one$}};
        \node[lc](n100) at (1.25,1.25) {\textcolor{white}{$\one\zero\zero$}};
        \node[lc](n101) at (3.75,1.25) {\textcolor{white}{$\one\zero\one$}};
        \node[lc](n110) at (1.25,3.75) {\textcolor{white}{$\one\one\zero$}};
        \node[pf](n111) at (3.75,3.75) {$\one\one\one$};
        \draw[thick,->] (n000) edge [loop left, distance=5mm] (n000);
        \draw[thick,->] (n001) edge (n100);
        \draw[thick,->] (n010) edge [bend left] (n001);
        \draw[thick,->] (n011) edge (n101);
        \draw[thick,->] (n100) edge (n010);
        \draw[thick,->] (n101) edge [bend left] (n110);
        \draw[thick,->] (n110) edge (n011);
        \draw[thick,->] (n111) edge [loop right, distance=5mm] (n111);
      \end{tikzpicture}}}
    \end{minipage}
    \quad\vrule\quad
    \begin{minipage}{.45\textwidth}
      \centerline{\scalebox{1}{\begin{tikzpicture}[>=to,auto]
        \tikzstyle{conf} = [rectangle, thick, draw]
        \tikzstyle{pf} = [rectangle, thick, draw, fill=black!20]
        \tikzstyle{lc} = [rectangle, thick, draw, fill=black!80]
        \node[lc](n000) at (0,0) {\textcolor{white}{$\zero\zero\zero$}};
        \node[lc](n001) at (2.5,0) {\textcolor{white}{$\zero\zero\one$}};
        \node[lc](n010) at (0,2.5) {\textcolor{white}{$\zero\one\zero$}};
        \node[lc](n011) at (2.5,2.5) {\textcolor{white}{$\zero\one\one$}};
        \node[lc](n100) at (1.25,1.25) {\textcolor{white}{$\one\zero\zero$}};
        \node[lc](n101) at (3.75,1.25) {\textcolor{white}{$\one\zero\one$}};
        \node[lc](n110) at (1.25,3.75) {\textcolor{white}{$\one\one\zero$}};
        \node[lc](n111) at (3.75,3.75) {\textcolor{white}{$\one\one\one$}};
        \draw[thick,->] (n000) edge (n100);
        \draw[thick,->] (n001) edge (n000);
        \draw[thick,->] (n010) edge [bend left=10] (n101);
        \draw[thick,->] (n011) edge (n001);
        \draw[thick,->] (n100) edge (n110);
        \draw[thick,->] (n101) edge [bend left=10] (n010);
        \draw[thick,->] (n110) edge (n111);
        \draw[thick,->] (n111) edge (n011);
      \end{tikzpicture}}}
    \end{minipage}
    \caption{Left panel: the parallel transition graph of positive canonical \BAC{} $\mathscr{C}_3^+$. Right panel: the parallel transition graph of negative canonical \BAC{} $\mathscr{C}_3^-$.}
    \label{fig:pcycles_tg}
  \end{center}
\end{figure}
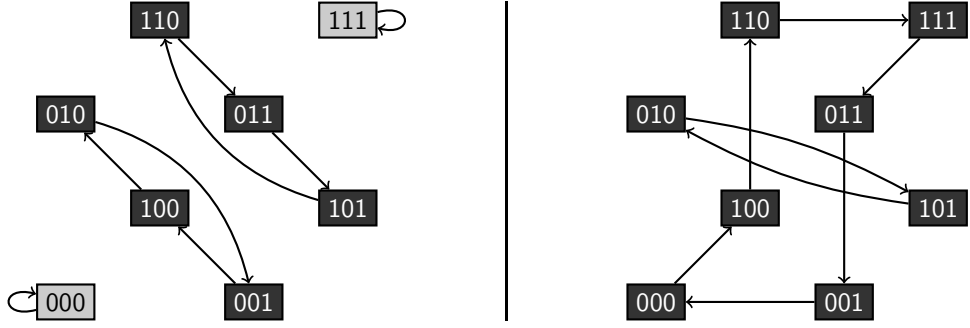

\begin{theorem}[\cite{J-Demongeot2012}] The order $\omega$, the numbers $\NBrec(p)$ and $\NBrecmin(p)$ of configurations of (minimal) period $p$, where $\NBrec(\omega)$ is the total number of recurring configurations, as well as the number $\NBatt(p)$ of attractors of period $p$ and the total number $\NBtotatt(\omega)$ of attractors of positive and negative \BACs{} are:\\[2mm]
  \centerline{
    \scalebox{0.88}{
      \setlength{\tabcolsep}{0.01ex} 
      \def\arraystretch{1.5}
      \begin{tabular}{|m{50mm}|m{55mm}|} 
        \begin{tabular}{m{50mm}} 
          \hline 
          \centering Positive cycles \\
          \centering $\mathscr{C}^+_n$
        \end{tabular}
        & \begin{tabular}{m{55mm}}
          \hline
          \centering Negative cycles \\
          \centering $\mathscr{C}^-_n$
        \end{tabular} \\[5pt]
        \hline\hline
        \begin{tabular}{m{50mm}} \centering $\omega = n$ \end{tabular} & 
        \begin{tabular}{m{55mm}} \centering $\omega = 2n$ \end{tabular} \\ 
        \hline
        \begin{tabular}{m{50mm}}
          \centering $\NBrecp(p) = 2^p$
        \end{tabular} &
        \begin{tabular}{m{55mm}}
          \centering $\NBrecn_n(p) = \neg(p|n) \cdot 2^{\frac{p}{2}}$
        \end{tabular}\\
        \hline
        \begin{tabular}{m{50mm}}
          \centering $\begin{array}{rcl} 
            \NBrecmin^+(p) & = & \underset{d|p}{\sum}\, \mu 
            \left( \frac{p}{d} \right) \cdot 2^{d}\\
            & = & \texttt{OEIS A27375}(p) 
          \end{array}$
        \end{tabular} &
        \begin{tabular}{m{55mm}}
          \centering $\begin{array}{rcl} 
            \NBrecmin^-_n(p) & = & \underset{k|p \text{ odd}}{\sum}\, 
            \mu(k) \cdot 2^{\frac{p}{2k}} 
          \end{array}$
        \end{tabular} \\
        \hline
        \begin{tabular}{m{50mm}} 
          \centering $\begin{array}{rcl} 
            \NBattp(p) & = & \frac{\NBrecmin^+(p)}{p} \\
            & = & \texttt{OEIS A1037}(p) 
          \end{array}$
        \end{tabular} &
        \begin{tabular}{m{55mm}} 
          \centering $\begin{array}{rcl} 
            \NBattn_n(p) & = & \frac{\NBrecmin^-_n(p)}{p} \\
            & = & \texttt{OEIS A48}(\frac{p}{2}) 
          \end{array}$ 
        \end{tabular}\\ 
        \hline
        \begin{tabular}{m{50mm}} 
          \centering $\begin{array}{rcl} 
            \NBtotattp(\omega) & = & \frac{1}{n}\, \underset{d|n}{\sum}\, 
            \euler \left( \frac{n}{d} \right) \cdot 2^{d} \\
            & = & \texttt{OEIS A31}(n) 
          \end{array}$ 
        \end{tabular} &
        \begin{tabular}{m{55mm}} 
          \centering $\begin{array}{rcl} 
            \NBtotattn(\omega) & = & \frac{1}{2n}\, \underset{k|2n 
            \text{ odd}}{\sum}\, \euler(k) \cdot 2^{\frac{n}{2k}} \\ 
            & = & \texttt{OEIS A16}(n) 
          \end{array}$ 
        \end{tabular}\\
        \hline
      \end{tabular}~\text{,}
    }
  }\smallskip
  
  \noindent where $\neg (p|n)$ is $0$ if $p$ divides $n$ and $1$ otherwise.
  \label{th:cycles_p}
\end{theorem}

Among the particularly interesting properties of a parallel \BAC{} that emerge from this theorem, it should be noted that all of its configurations are recurring, as illustrated in Figure~\ref{fig:pcycles_tg}, and that the total number of attractors, for a given period or not, is exponential according to its size $n$. 
Moreover, as a corollary, for the two types of cycles, the order is reached, and positive cycles admit two stable configurations $x$ and $\bar{x}^V$ (where $x$ is the configuration in which all the automata are at state $\zero$ for canonical \BACs{}). 
Finally, another important point is that this combinatorial study induces the complete characterisation of the structure of the parallel transition graphs of both positive and negative cycles.\smallskip

It is also interesting to notice that these results go beyond the parallel updating mode and extend to the block-sequential updating modes. 
Block-sequential updating modes were introduced by Robert~\cite{J-Robert1969}. 
They are deterministic periodic updating modes defined by ordered partitions of $V$. 
Given a period $p$, such an updating mode can be defined by a function $\mu: V \to 
\NN/p\NN$. 
Indeed, in~\cite{C-Goles2010}, the authors showed that the dynamics of a \BAC{} of size $n$ and sign $s \in {+, -}$ evolving according to a block-sequential updating mode is in essence  equivalent to that of a \BAC{} of smaller size and same sign evolving in parallel. 
The proof rests on substitutions of local transition functions according to their  execution over time.
Hence, to understand the dynamics of a \BAC{} evolving according to a block-sequential updating mode is a matter of understanding the dynamics in parallel of a smaller \BAC{}.

\subsection{Asynchronous \BACs{}}
\label{sec:BACs:a}

\begin{figure}[t!]
  \begin{center}
    \begin{minipage}{.45\textwidth}
      \centerline{\scalebox{1}{\begin{tikzpicture}[>=to,auto]
        \tikzstyle{conf} = [rectangle, draw]
        \tikzstyle{pf} = [rectangle, thick, draw, fill=black!20]
        \tikzstyle{lc} = [rectangle, thick, draw, fill=black!80]
        \node[pf](n000) at (0,0) {$\zero\zero\zero$};
        \node[conf](n001) at (2.5,0) {$\zero\zero\one$};
        \node[conf](n010) at (0,2.5) {$\zero\one\zero$};
        \node[conf](n011) at (2.5,2.5) {$\zero\one\one$};
        \node[conf](n100) at (1.25,1.25) {$\one\zero\zero$};
        \node[conf](n101) at (3.75,1.25) {$\one\zero\one$};
        \node[conf](n110) at (1.25,3.75) {$\one\one\zero$};
        \node[pf](n111) at (3.75,3.75) {$\one\one\one$};
        \draw[thick,Green4,->] (n000) edge [loop left, distance=5mm] (n000);
        \draw[thick,Blue3,->] (n000) edge [loop left, distance=7mm] (n000);
        \draw[thick,Red3,->] (n000) edge [loop left, distance=9mm] (n000);
        \draw[thick,Green4,->] (n001) edge (n000);
        \draw[thick,Blue3,->] (n001) edge [loop right, distance=5mm] (n001);
        \draw[thick,Red3,->] (n001) edge (n101);        
        \draw[thick,Green4,->] (n010) edge (n011);
        \draw[thick,Blue3,->] (n010) edge (n000);
        \draw[thick,Red3,->] (n010) edge [loop left, distance=5mm] (n010);
        \draw[thick,Green4,->] (n011) edge [loop right, distance=5mm] (n011);
        \draw[thick,Blue3,->] (n011) edge (n001);
        \draw[thick,Red3,->] (n011) edge (n111);
        \draw[thick,Green4,->] (n100) edge [loop left, distance=5mm] (n100);
        \draw[thick,Blue3,->] (n100) edge (n110);
        \draw[thick,Red3,->] (n100) edge (n000);
        \draw[thick,Green4,->] (n101) edge (n100);
        \draw[thick,Blue3,->] (n101) edge (n111);
        \draw[thick,Red3,->] (n101) edge [loop right, distance=5mm] (n101);
        \draw[thick,Green4,->] (n110) edge (n111);
        \draw[thick,Blue3,->] (n110) edge [loop left, distance=5mm] (n110);
        \draw[thick,Red3,->] (n110) edge (n010);
        \draw[thick,Green4,->] (n111) edge [loop right, distance=5mm] (n111);
        \draw[thick,Blue3,->] (n111) edge [loop right, distance=7mm] (n111);
        \draw[thick,Red3,->] (n111) edge [loop right, distance=9mm] (n111);
      \end{tikzpicture}}}
    \end{minipage}
    \quad\vrule\quad
    \begin{minipage}{.45\textwidth}
      \centerline{\scalebox{1}{\begin{tikzpicture}[>=to,auto]
        \tikzstyle{conf} = [rectangle, draw]
        \tikzstyle{pf} = [rectangle, thick, draw, fill=black!20]
        \tikzstyle{lc} = [rectangle, thick, draw, fill=black!80]
        \node[lc](n000) at (0,0) {\textcolor{white}{$\zero\zero\zero$}};
        \node[lc](n001) at (2.5,0) {\textcolor{white}{$\zero\zero\one$}};
        \node[conf](n010) at (0,2.5) {$\zero\one\zero$};
        \node[lc](n011) at (2.5,2.5) {\textcolor{white}{$\zero\one\one$}};
        \node[lc](n100) at (1.25,1.25) {\textcolor{white}{$\one\zero\zero$}};
        \node[conf](n101) at (3.75,1.25) {$\one\zero\one$};
        \node[lc](n110) at (1.25,3.75) {\textcolor{white}{$\one\one\zero$}};
        \node[lc](n111) at (3.75,3.75) {\textcolor{white}{$\one\one\one$}};
        \draw[thick,Green4,->] (n000) edge [loop left, distance=5mm] (n000);
        \draw[thick,Blue3,->] (n000) edge [loop left, distance=7mm] (n000);
        \draw[thick,Red3,->] (n000) edge (n100);
        \draw[thick,Green4,->] (n001) edge (n000);
        \draw[thick,Blue3,->] (n001) edge [loop right, distance=5mm] (n001);
        \draw[thick,Red3,->] (n001) edge [loop right, distance=7mm] (n001);
        \draw[thick,Green4,->] (n010) edge (n011);
        \draw[thick,Blue3,->] (n010) edge (n000);
        \draw[thick,Red3,->] (n010) edge (n110);
        \draw[thick,Green4,->] (n011) edge [loop right, distance=5mm] (n011);
        \draw[thick,Blue3,->] (n011) edge (n001);
        \draw[thick,Red3,->] (n011) edge [loop right, distance=7mm] (n011);
        \draw[thick,Green4,->] (n100) edge [loop left, distance=5mm] (n100);
        \draw[thick,Blue3,->] (n100) edge (n110);
        \draw[thick,Red3,->] (n100) edge [loop left, distance=7mm] (n100);
        \draw[thick,Green4,->] (n101) edge (n100);
        \draw[thick,Blue3,->] (n101) edge (n111);
        \draw[thick,Red3,->] (n101) edge (n001);
        \draw[thick,Green4,->] (n110) edge (n111);
        \draw[thick,Blue3,->] (n110) edge [loop left, distance=5mm] (n110);
        \draw[thick,Red3,->] (n110) edge [loop left, distance=7mm] (n110);
        \draw[thick,Green4,->] (n111) edge [loop right, distance=5mm] (n111);
        \draw[thick,Blue3,->] (n111) edge [loop right, distance=7mm] (n111);
        \draw[thick,Red3,->] (n111) edge (n011);
      \end{tikzpicture}}}
    \end{minipage}
    \caption{Left panel: asynchronous transition graph of positive canonical \BAC{} $\mathscr{C}_3^+$. 
      Right panel: asynchronous transition graph of negative canonical \BAC{} $\mathscr{C}_3^-$. 
      In both graphs, every $\textcolor{Red3}{\to}$ (resp. $\textcolor{Blue3}{\to}$ and $\textcolor{Green4}{\to}$) represents an update of automaton $0$ (resp. $1$ and $2$).}
    \label{fig:acycles_tg}
  \end{center}
\end{figure}
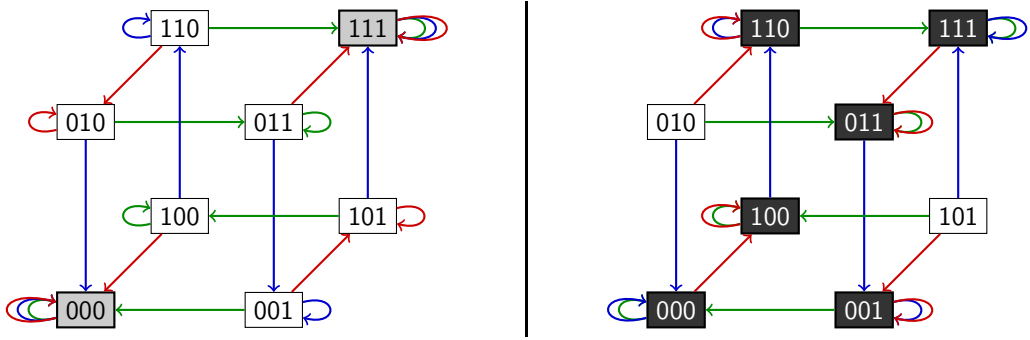

In combinatorial terms, the dynamics of asynchronous \BACs{} are much simpler than that of parallel \BACs{}.
It is in~\cite{J-Remy2003} that we find the first characterisation of the attractors of asynchronous \BACs{}.
The general idea of these results is based on the concept of instability. In a given configuration of the network, an automaton is said to be unstable if the application of its local transition function would then make its state change. A 
configuration is unstable when it has at least one unstable automaton. 
In particular, the authors showed that asynchronism makes it possible to reduce the number of instabilities, until there are none (resp. only one) left in the case of positive (resp. negative) \BACs{}. This  property  implies the existence of at least one stable configuration in the positive case. In the negative case, it implies  the absence of stable configurations and thereby the existence of at least one stable oscillation. 
Theorem~\ref{th:cycles_a} derives almost directly from this work. It is illustrated by Figure~\ref{fig:acycles_tg}.

\begin{theorem}[\cite{J-Remy2003}] A positive \BAC{} $\mathscr{C}_n^+$ has two attractors which are two stable configurations $x$ and $\bar{x}^V$. 
  A negative cycle $\mathscr{C}_n^-$ has a single attractor of length $2n$.
  \label{th:cycles_a}
\end{theorem}

\begin{figure}[t!]
  \begin{center}
    \begin{minipage}{.45\textwidth}
      \centerline{\scalebox{1}{\begin{tikzpicture}[>=latex,auto]
        \tikzstyle{type} = []
        \tikzstyle{node} = [circle, thick, draw]
        \node[type] (d) at (-1,1.5) {$\mathscr{D}_{2,2}^{+,+}$};
        \node[node] (n0) at (0,0) {$0$};
        \node[node] (n1) at (1.5,1.5) {$1$};
        \node[node] (n2) at (3,0) {$2$};
        \draw[thick, ->] (n0) edge [bend left] node {$+$} (n1);
        \draw[thick, ->] (n1) edge [bend left] node {$+$} (n0);
        \draw[thick, ->] (n1) edge [bend right] node [swap] {$+$} (n2);
        \draw[thick, ->] (n2) edge [bend right] node [swap] {$+$} (n1);
      \end{tikzpicture}}}
    \end{minipage}
    \quad\vrule\quad
    \begin{minipage}{.45\textwidth}
      \centerline{\scalebox{1}{\begin{tikzpicture}[>=latex,auto]
        \tikzstyle{type} = []
        \tikzstyle{node} = [circle, thick, draw]
        \node[type] (d) at (-1,1.5) {$\mathscr{D}_{1,3}^{+,+}$};
        \node[node] (n0) at (0.5,.75) {$0$};
        \node[node] (n1) at (2,1.5) {$1$};
        \node[node] (n2) at (2,0) {$2$};
        \draw[thick, ->] (n0) edge [loop left] node {$+$} (n0);
        \draw[thick, ->] (n0) edge [bend left] node {$+$} (n1);
        \draw[thick, ->] (n1) edge [bend left] node {$+$} (n2);
        \draw[thick, ->] (n2) edge [bend left] node {$+$} (n0);
      \end{tikzpicture}}}
    \end{minipage}\smallskip
    
    \hrule\smallskip
    
    \begin{minipage}{.45\textwidth}
      \centerline{\scalebox{1}{\begin{tikzpicture}[>=latex,auto]
        \tikzstyle{type} = []
        \tikzstyle{node} = [circle, thick, draw]
        \node[type] (d) at (-1,1.5) {$\mathscr{D}_{2,2}^{-,+}$};
        \node[node] (n0) at (0,0) {$0$};
        \node[node] (n1) at (1.5,1.5) {$1$};
        \node[node] (n2) at (3,0) {$2$};
        \draw[thick, ->] (n0) edge [bend left] node {$-$} (n1);
        \draw[thick, ->] (n1) edge [bend left] node {$+$} (n0);
        \draw[thick, ->] (n1) edge [bend right] node [swap] {$+$} (n2);
        \draw[thick, ->] (n2) edge [bend right] node [swap] {$+$} (n1);
      \end{tikzpicture}}}
    \end{minipage}
    \quad\vrule\quad
    \begin{minipage}{.45\textwidth}
      \centerline{\scalebox{1}{\begin{tikzpicture}[>=latex,auto]
        \tikzstyle{type} = []
        \tikzstyle{node} = [circle, thick, draw]
        \node[type] (d) at (-1,1.5) {$\mathscr{D}_{1,3}^{-,+}$};
        \node[node] (n0) at (0.5,.75) {$0$};
        \node[node] (n1) at (2,1.5) {$1$};
        \node[node] (n2) at (2,0) {$2$};
        \draw[thick, ->] (n0) edge [loop left] node {$-$} (n0);
        \draw[thick, ->] (n0) edge [bend left] node {$+$} (n1);
        \draw[thick, ->] (n1) edge [bend left] node {$+$} (n2);
        \draw[thick, ->] (n2) edge [bend left] node {$+$} (n0);
      \end{tikzpicture}}}
    \end{minipage}\smallskip

    \hrule\smallskip
    
    \begin{minipage}{.45\textwidth}
      \centerline{\scalebox{1}{\begin{tikzpicture}[>=latex,auto]
        \tikzstyle{type} = []
        \tikzstyle{node} = [circle, thick, draw]
        \node[type] (d) at (-1,1.5) {$\mathscr{D}_{2,2}^{-,-}$};
        \node[node] (n0) at (0,0) {$0$};
        \node[node] (n1) at (1.5,1.5) {$1$};
        \node[node] (n2) at (3,0) {$2$};
        \draw[thick, ->] (n0) edge [bend left] node {$-$} (n1);
        \draw[thick, ->] (n1) edge [bend left] node {$+$} (n0);
        \draw[thick, ->] (n1) edge [bend right] node [swap] {$+$} (n2);
        \draw[thick, ->] (n2) edge [bend right] node [swap] {$-$} (n1);
      \end{tikzpicture}}}
    \end{minipage}
    \quad\vrule\quad
    \begin{minipage}{.45\textwidth}
      \centerline{\scalebox{1}{\begin{tikzpicture}[>=latex,auto]
        \tikzstyle{type} = []
        \tikzstyle{node} = [circle, thick, draw]
        \node[type] (d) at (-1,1.5) {$\mathscr{D}_{1,3}^{-,-}$};
        \node[node] (n0) at (0.5,.75) {$0$};
        \node[node] (n1) at (2,1.5) {$1$};
        \node[node] (n2) at (2,0) {$2$};
        \draw[thick, ->] (n0) edge [loop left] node {$-$} (n0);
        \draw[thick, ->] (n0) edge [bend left] node {$+$} (n1);
        \draw[thick, ->] (n1) edge [bend left] node {$+$} (n2);
        \draw[thick, ->] (n2) edge [bend left] node {$-$} (n0);
      \end{tikzpicture}}}
    \end{minipage}\smallskip
  \caption{Top panel: Interaction graphs of two positive canonical \BADCs{}.
    Middle panel: Interaction graphs of two mixed canonical \BADCs{}. 
    Bottom panel: Interaction graphs of two negative canonical \BADCs{}.
    These six networks will serve as examples in the sequel.}
  \label{fig:pdcycles_ig}
  \end{center}
\end{figure}
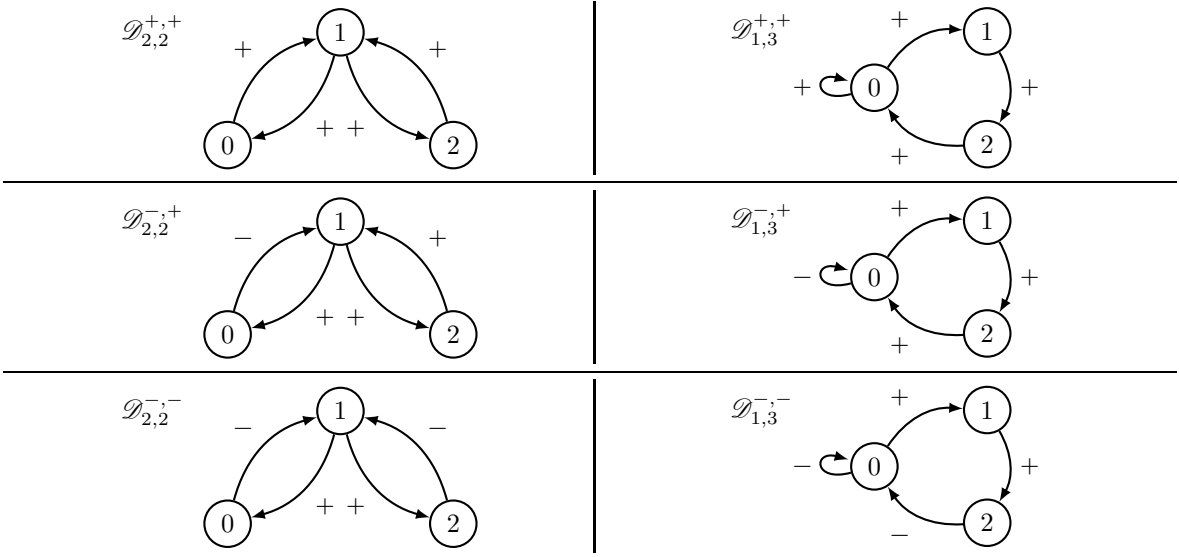

It is easy to see that the results stated in this theorem are strongly related to 
those of Theorem~\ref{th:cycles_p}. 
Indeed, first, the two stable configurations of asynchronous positive cycles are identical to those of these same cycles in parallel.
And second, the unique stable oscillation of length $2n$ of a negative cycle $\mathscr{C}_n^-$ under the asynchronous mode is identical to that of period $\omega$ of parallel negative cycles. 
The two configurations $x$ and $\bar{x}^V$ which are  stable configurations in the case of positive cycles,  belong to the stable oscillations in the case of negative cycles.\smallskip

Finally, notice that this approach based on instabilities has been generalised in~\cite{T-Noual2012,T-Sene2012} to show the validity of Theorem~\ref{th:thomas_cp} regardless of the updating mode (see Section~\ref{sec:semres_cycles}).

\section{\BADC{} dynamics}
\label{sec:BADCs}

Now that the structural and combinatorial properties of the parallel and asynchronous transition graphs of the \BACs{} are established, we present in this section those related to \BADCs{}.

\subsection{Parallel \BADCs{}}
\label{sec:BADCs:p}

In this section, we focus on \BADCs{} evolving according to the parallel updating. 
The method used to obtain the results presented follows the lines of  method  used for  \BACs{}.

\subsubsection{Definitions and notations}

Besides the four quantities $\NBrec(p)$, $\NBrecmin(p)$, $\NBatt(p)$ and $\NBtotatt(\omega)$ that we are going to use, we introduce here other definitions and notations to characterise the dynamics of \BADCs{} that are \BACs{} tangentially interconnected. 
In this sense, given any \BADC{} $\mathscr{D}_{\ell, r}^s$, where $s \in \{(+, +), (-, +), (-, -)\}$, two quantities defined by means of $\ell$ and $r$ will be particularly useful, $\Delta = \gcd(\ell, r)$ and $\Delta_p = \gcd(\Delta, p)$.\smallskip

Moreover, the results call on combinatorics on words.
So we introduce the Lucas' and Perrin's sequences. These sequences  will allow us to count the number of recurring configurations. 
The Lucas' sequence $(\lucas(n))_{n \in \NN^\ast}$ (\texttt{OEIS A204}) is defined by $\lucas(1) = 1$, $\lucas(2) = 3$ and for all $n > 2$, $\lucas(n) = \lucas(n-1) + \lucas(n-2)$, and counts the number of binary necklaces of size $n$ without the factor $\zero\zero$. 
The Perrin's sequence $(\perrin(n))_{n \in \NN}$ (\texttt{OEIS A1608}) is defined by $\perrin(0) = 3$, $\perrin(1) = 0$, $\perrin(2) = 2$ and for all $n > 2$, $\perrin(n) = \perrin(n-2) + \perrin(n-3)$, and counts the number of binary necklaces of size $n$ without the factors $\zero\zero$ and $\one\one\one$.

\subsubsection{Results}

On the basis of the previous definitions and notations, Theorem~\ref{th:dcycles_p} below gives the qualitative characterisation of the dynamics of the different types of \BADCs{}. 
It will be illustrated on six distinct \BADCs{} depicted in Figure~\ref{fig:pdcycles_ig}. 
The details of the proofs can be found in~\cite{C-Noual2012,T-Noual2012,T-Sene2012}. 

\begin{theorem}[\cite{C-Noual2012}] The order $\omega$, the numbers $\NBrec(p)$ and $\NBrecmin(p)$ of configurations of (minimal) period $p$, where $\NBrec(\omega)$ represents the total number of recurring configurations, as well as the number $\NBatt(p)$ of asymptotic behaviours of period $p$, and the total number $\NBtotatt(\omega)$ asymptotic behaviours of the positive, mixed and negative \BADCs{} are:\\[2mm]
  \centerline{
    \scalebox{0.88}{
      \setlength{\tabcolsep}{0.01ex}
      \def\arraystretch{1.5}
      \begin{tabular}{|m{25mm}|m{60mm}|m{60mm}|}
        \begin{tabular}{m{25mm}} 
          \hline 
          \centering Positive \\
          \centering double-cycles \\
          \centering $\mathscr{D}^{+,+}_{\ell,r}$
        \end{tabular}
        & \begin{tabular}{m{60mm}}
          \hline
          \centering Mixed \\
          \centering double-cycles \\
          \centering $\mathscr{D}^{-,+}_{\ell,r}$
          \end{tabular}
        & \begin{tabular}{m{60mm}}
          \hline
          \centering Negative \\
          \centering double-cycles \\
          \centering $\mathscr{D}^{-,-}_{\ell,r}$
          \end{tabular} \\[5pt]
        \hline\hline
        \begin{tabular}{m{25mm}} \centering $\omega = \Delta$ \end{tabular} & 
        \begin{tabular}{m{60mm}} \centering $\omega = r$ \end{tabular} & 
        \begin{tabular}{m{60mm}} \centering $\begin{cases}
            \frac{\ell+r}{2} & \text{if } \frac{\ell+r}{\Delta} = 4 \\
            \ell + r & \text{sinon}
          \end{cases}$
        \end{tabular} \\ 
        \hline
        \begin{tabular}{m{25mm}}
        \centering $\NBrecp(p)$
        \end{tabular} &				
        \begin{tabular}{m{60mm}}
          \centering $\texttt{X}^{-,+}_{\ell}(p) = \neg (p|\ell) \cdot 
          \lucas(\frac{p}{\Delta_p})^{\Delta_p}$
        \end{tabular} &
        \begin{tabular}{m{60mm}}
          \centering $\texttt{X}^{-,-}_{\Delta}(p) = \neg (p|\Delta) \cdot 
          \perrin(\frac{p}{\Delta_p})^{\Delta_p}$
        \end{tabular}\\
        \hline
        \begin{tabular}{m{25mm}}
          \centering $\NBrecmin^+(p)$
        \end{tabular} &
        \begin{tabular}{m{60mm}}
          \centering $\NBrecmin^{-,+}_{\ell}(p) = 
          \underset{\substack{d|p\\\neg (d|\ell)}}{\sum}\, \mu \left( 
          \frac{p}{d} \right) \cdot 
          \lucas(\frac{d}{\Delta_d})^{\Delta_d}$
        \end{tabular} &
        \begin{tabular}{m{60mm}}
          \centering $\NBrecmin^{-,-}_\Delta(p) = 
          \underset{\substack{d|p\\\neg (d|\Delta)}}{\sum}\, \mu \left( 
          \frac{p}{d} \right) \cdot 
          \perrin(\frac{d}{\Delta_d})^{\Delta_d}$
        \end{tabular}\\
        \hline
        \begin{tabular}{m{25mm}} 
          \centering $\NBattp(p)$
        \end{tabular} &
        \begin{tabular}{m{60mm}} 
          \centering $\NBatt^{-,+}_\ell (p) = 
          \frac{\NBrecmin^{-,+}_{\ell}(p)}{p}$
        \end{tabular} &
        \begin{tabular}{m{60mm}} 
          \centering $\NBatt^{-,-}_\Delta (p) = 
          \frac{\NBrecmin^{-,-}_{\Delta}(p)}{p}$ 
        \end{tabular}\\ 
        \hline
        \begin{tabular}{m{25mm}} 
          \centering $\NBtotattp(\omega)$ 
        \end{tabular} &
        \begin{tabular}{m{60mm}} 
          \centering $\NBtotatt^{-,+}_\ell (\omega) = \frac{1}{r} 
          \underset{\substack{d|r\\ \neg (d|\ell)}}{\sum}\, \euler \left( 
          \frac{r}{d} \right) \cdot 
          \lucas(\frac{d}{\Delta_d})^{\Delta_d}$ 
        \end{tabular} &
        \begin{tabular}{m{60mm}} 
          \centering $\NBtotatt^{-,-}_\Delta (\omega) = \frac{1}{n} 
          \underset{\substack{d|n\\ \neg (d|\Delta)}}{\sum}\, \euler \left( 
          \frac{n}{d} \right) \cdot 
          \perrin(\frac{d}{\Delta_d})^{\Delta_d}$ 
        \end{tabular}\\
        \hline
      \end{tabular}~\text{,}
    }
  }\smallskip
  
  \noindent where $\neg (p | m) $ equals $0$ if $p$ divides $m$ and $1$ otherwise.
  \label{th:dcycles_p}
\end{theorem}

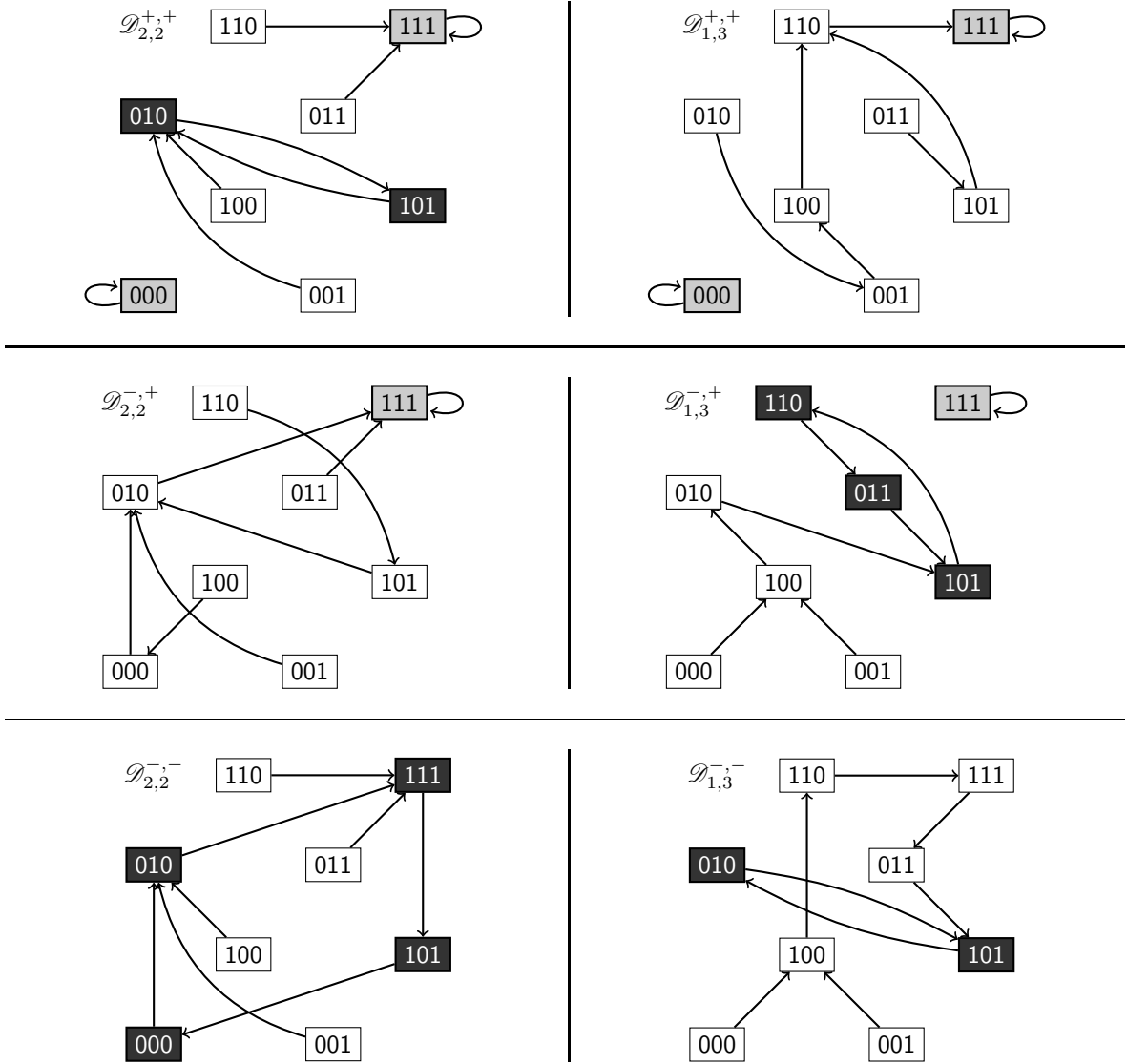
\begin{figure}[t!]
  \begin{center}
    \begin{minipage}{.45\textwidth}
      \centerline{\scalebox{1}{\begin{tikzpicture}[>=to,auto]
        \tikzstyle{type} = []
        \tikzstyle{conf} = [rectangle, draw]
        \tikzstyle{pf} = [rectangle, thick, draw, fill=black!20]
        \tikzstyle{lc} = [rectangle, thick, draw, fill=black!80]
        \node[type] (d) at (0,3.75) {$\mathscr{D}_{2,2}^{+,+}$};
        \node[pf](n000) at (0,0) {$\zero\zero\zero$};
        \node[conf](n001) at (2.5,0) {$\zero\zero\one$};
        \node[lc](n010) at (0,2.5) {\textcolor{white}{$\zero\one\zero$}};
        \node[conf](n011) at (2.5,2.5) {$\zero\one\one$};
        \node[conf](n100) at (1.25,1.25) {$\one\zero\zero$};
        \node[lc](n101) at (3.75,1.25) {\textcolor{white}{$\one\zero\one$}};
        \node[conf](n110) at (1.25,3.75) {$\one\one\zero$};
        \node[pf](n111) at (3.75,3.75) {$\one\one\one$};
        \draw[thick,->] (n000) edge [loop left] (n000);
        \draw[thick,->] (n001) edge [bend left] (n010);
        \draw[thick,->] (n010) edge [bend left=10] (n101);
        \draw[thick,->] (n011) edge (n111);
        \draw[thick,->] (n100) edge (n010);
        \draw[thick,->] (n101) edge [bend left=10] (n010);
        \draw[thick,->] (n110) edge (n111);
        \draw[thick,->] (n111) edge [loop right] (n111);
      \end{tikzpicture}}}
    \end{minipage}
    \quad\vrule\quad
    \begin{minipage}{.45\textwidth}
      \centerline{\scalebox{1}{\begin{tikzpicture}[>=to,auto]
        \tikzstyle{type} = []
        \tikzstyle{conf} = [rectangle, draw]
        \tikzstyle{pf} = [rectangle, thick, draw, fill=black!20]
        \tikzstyle{lc} = [rectangle, thick, draw, fill=black!80]
        \node[type] (d) at (0,3.75) {$\mathscr{D}_{1,3}^{+,+}$};
        \node[pf](n000) at (0,0) {$\zero\zero\zero$};
        \node[conf](n001) at (2.5,0) {$\zero\zero\one$};
        \node[conf](n010) at (0,2.5) {$\zero\one\zero$};
        \node[conf](n011) at (2.5,2.5) {$\zero\one\one$};
        \node[conf](n100) at (1.25,1.25) {$\one\zero\zero$};
        \node[conf](n101) at (3.75,1.25) {$\one\zero\one$};
        \node[conf](n110) at (1.25,3.75) {$\one\one\zero$};
        \node[pf](n111) at (3.75,3.75) {$\one\one\one$};
        \draw[thick,->] (n000) edge [loop left] (n000);
        \draw[thick,->] (n001) edge (n100);
        \draw[thick,->] (n010) edge [bend right] (n001);
        \draw[thick,->] (n011) edge (n101);
        \draw[thick,->] (n100) edge (n110);
        \draw[thick,->] (n101) edge [bend right] (n110);
        \draw[thick,->] (n110) edge (n111);
        \draw[thick,->] (n111) edge [loop right] (n111);
      \end{tikzpicture}}}
    \end{minipage}\bigskip
    
    \hrule\bigskip
    
    \begin{minipage}{.45\textwidth}
      \centerline{\scalebox{1}{\begin{tikzpicture}[>=to,auto]
        \tikzstyle{type} = []
        \tikzstyle{conf} = [rectangle, draw]
        \tikzstyle{pf} = [rectangle, thick, draw, fill=black!20]
        \tikzstyle{lc} = [rectangle, thick, draw, fill=black!80]
        \node[type] (d) at (0,3.75) {$\mathscr{D}_{2,2}^{-,+}$};
        \node[conf](n000) at (0,0) {$\zero\zero\zero$};
        \node[conf](n001) at (2.5,0) {$\zero\zero\one$};
        \node[conf](n010) at (0,2.5) {$\zero\one\zero$};
        \node[conf](n011) at (2.5,2.5) {$\zero\one\one$};
        \node[conf](n100) at (1.25,1.25) {$\one\zero\zero$};
        \node[conf](n101) at (3.75,1.25) {$\one\zero\one$};
        \node[conf](n110) at (1.25,3.75) {$\one\one\zero$};
        \node[pf](n111) at (3.75,3.75) {$\one\one\one$};
        \draw[thick,->] (n000) edge (n010);
        \draw[thick,->] (n001) edge [bend left] (n010);
        \draw[thick,->] (n010) edge (n111);
        \draw[thick,->] (n011) edge (n111);
        \draw[thick,->] (n100) edge (n000);
        \draw[thick,->] (n101) edge (n010);
        \draw[thick,->] (n110) edge [bend left] (n101);
        \draw[thick,->] (n111) edge [loop right] (n111);
      \end{tikzpicture}}}
    \end{minipage}
    \quad\vrule\quad
    \begin{minipage}{.45\textwidth}
      \centerline{\scalebox{1}{\begin{tikzpicture}[>=to,auto]
        \tikzstyle{type} = []
        \tikzstyle{conf} = [rectangle, draw]
        \tikzstyle{pf} = [rectangle, thick, draw, fill=black!20]
        \tikzstyle{lc} = [rectangle, thick, draw, fill=black!80]
        \node[type] (d) at (0,3.75) {$\mathscr{D}_{1,3}^{-,+}$};
        \node[conf](n000) at (0,0) {$\zero\zero\zero$};
        \node[conf](n001) at (2.5,0) {$\zero\zero\one$};
        \node[conf](n010) at (0,2.5) {$\zero\one\zero$};
        \node[lc](n011) at (2.5,2.5) {\textcolor{white}{$\zero\one\one$}};
        \node[conf](n100) at (1.25,1.25) {$\one\zero\zero$};
        \node[lc](n101) at (3.75,1.25) {\textcolor{white}{$\one\zero\one$}};
        \node[lc](n110) at (1.25,3.75) {\textcolor{white}{$\one\one\zero$}};
        \node[pf](n111) at (3.75,3.75) {$\one\one\one$};
        \draw[thick,->] (n000) edge (n100);
        \draw[thick,->] (n001) edge (n100);
        \draw[thick,->] (n010) edge (n101);
        \draw[thick,->] (n011) edge (n101);
        \draw[thick,->] (n100) edge (n010);
        \draw[thick,->] (n101) edge [bend right] (n110);
        \draw[thick,->] (n110) edge (n011);
        \draw[thick,->] (n111) edge [loop right] (n111);
      \end{tikzpicture}}}
    \end{minipage}\bigskip

    \hrule\bigskip
    
    \begin{minipage}{.45\textwidth}
      \centerline{\scalebox{1}{\begin{tikzpicture}[>=to,auto]
        \tikzstyle{type} = []
        \tikzstyle{conf} = [rectangle, draw]
        \tikzstyle{pf} = [rectangle, thick, draw, fill=black!20]
        \tikzstyle{lc} = [rectangle, thick, draw, fill=black!80]
        \node[type] (d) at (0,3.75) {$\mathscr{D}_{2,2}^{-,-}$};
        \node[lc](n000) at (0,0) {\textcolor{white}{$\zero\zero\zero$}};
        \node[conf](n001) at (2.5,0) {$\zero\zero\one$};
        \node[lc](n010) at (0,2.5) {\textcolor{white}{$\zero\one\zero$}};
        \node[conf](n011) at (2.5,2.5) {$\zero\one\one$};
        \node[conf](n100) at (1.25,1.25) {$\one\zero\zero$};
        \node[lc](n101) at (3.75,1.25) {\textcolor{white}{$\one\zero\one$}};
        \node[conf](n110) at (1.25,3.75) {$\one\one\zero$};
        \node[lc](n111) at (3.75,3.75) {\textcolor{white}{$\one\one\one$}};
        \draw[thick,->] (n000) edge (n010);
        \draw[thick,->] (n001) edge [bend left] (n010);
        \draw[thick,->] (n010) edge (n111);
        \draw[thick,->] (n011) edge (n111);
        \draw[thick,->] (n100) edge (n010);
        \draw[thick,->] (n101) edge (n000);
        \draw[thick,->] (n110) edge (n111);
        \draw[thick,->] (n111) edge (n101);
      \end{tikzpicture}}}
    \end{minipage}
    \quad\vrule\quad
    \begin{minipage}{.45\textwidth}
      \centerline{\scalebox{1}{\begin{tikzpicture}[>=to,auto]
        \tikzstyle{type} = []
        \tikzstyle{conf} = [rectangle, draw]
        \tikzstyle{pf} = [rectangle, thick, draw, fill=black!20]
        \tikzstyle{lc} = [rectangle, thick, draw, fill=black!80]
        \node[type] (d) at (0,3.75) {$\mathscr{D}_{1,3}^{-,-}$};
        \node[conf](n000) at (0,0) {$\zero\zero\zero$};
        \node[conf](n001) at (2.5,0) {$\zero\zero\one$};
        \node[lc](n010) at (0,2.5) {\textcolor{white}{$\zero\one\zero$}};
        \node[conf](n011) at (2.5,2.5) {$\zero\one\one$};
        \node[conf](n100) at (1.25,1.25) {$\one\zero\zero$};
        \node[lc](n101) at (3.75,1.25) {\textcolor{white}{$\one\zero\one$}};
        \node[conf](n110) at (1.25,3.75) {$\one\one\zero$};
        \node[conf](n111) at (3.75,3.75) {$\one\one\one$};
        \draw[thick,->] (n000) edge (n100);
        \draw[thick,->] (n001) edge (n100);
        \draw[thick,->] (n010) edge [bend left=10] (n101);
        \draw[thick,->] (n011) edge (n101);
        \draw[thick,->] (n100) edge (n110);
        \draw[thick,->] (n101) edge [bend left=10] (n010);
        \draw[thick,->] (n110) edge (n111);
        \draw[thick,->] (n111) edge (n011);
      \end{tikzpicture}}}
    \end{minipage}\smallskip
  \caption{Parallel transition graphs of the canonical \BADCs{} depicted in Figure~\ref{fig:pdcycles_ig}, where $\diamond = \lor$.}
  \label{fig:pdcycles_tg}
  \end{center}
\end{figure}

Among other things (see Figure~\ref{fig:pdcycles_ig}), it emerges from these results that \emph{(i)} positive \BADCs{} have two stable configurations $x$ and $\bar{x}^V$ (with $x = (\zero, ..., \zero)$ when they are canonical) and that they have an asymptotic behaviour similar to positive \BACs{} of the same order, \emph{(ii)} mixed \BADCs{} have a single stable configuration, and \emph{(iii)} negative \BADCs{} have no stable configurations.\smallskip

Also, we remark that, unlike \BACs{}, the order of \BADCs{} is not necessarily reached. 
Finally, this theorem highlights again that, like \BACs{}, \BADCs{} admit an exponential number of attractors according to their size. 
However, despite its exponential nature, the number of \BADC{} attractors is significantly smaller than that of \BACs{}. 
In other words, the intersections of cycles seem to participate strongly in the reduction of asymptotic degrees of freedom of interaction networks. 
Based on this idea, studies have been conducted to compare $\NBtotatt^+ (n)$ and $\NBtotatt^- (2n)$ of \BACs{} with quantities $\NBtotatt^+ (\Delta)$, $\NBtotatt^{-,+}_\ell (r)$ and $\NBtotatt^{-,-}_\Delta (\ell+r)$ of \BADCs{}.\smallskip

Consider a network $f$ of order $\omega$, so that $f = \mathscr{C}_n^s$ or $f = \mathscr{D}_{\ ell, r}^{s, s'}$, where $s, s' \in \{+, -\}$. 
Let $p$ be a divisor of $\omega$. 
Also note $\quant \in \{\NBrec, \NBrecmin, \NBatt, \NBtotatt\}$ one of the four quantities analysed. 
It has been demonstrated in~\cite{T-Noual2012,T-Sene2012} that $\quant^{+, +}_{\ell, r}(p) = \quant^+(p)$, $\NBrecnp_{\ell, r}(p) \leq \NBrecp(p)$, $\NBrecnn_{\ ell, r}(p) \leq \NBrecp(p)$ and $\quant_{\ell, \ell}^{s, s}(p) = \quant^s_\ell(p)$.\smallskip

However, the number of recurring configurations of a \BADC{}, $\NBrec_{\ell,r}^{s,s}(p)$ has been bounded more finely as a function of $\NBrec_n^+(p)$ and $\NBrec_n^-(p)$, based on the previous results as well as on the relation of the Lucas sequence with the golden ratio~\cite{L-Ribenboim1996}, denoted by $\gold = (1+\sqrt{5})/2$ (root of $x^2 - x - 1 = 0$) and that of the Perrin sequence with the plastic number~\cite{L-VanderLaan1960}, denoted by $\plastic = \sqrt[3]{\frac{1}{2} + \frac{1}{6}\sqrt{\frac{23}{3}}} + \sqrt[3]{\frac{1}{2} - \frac{1}{6}\sqrt{\frac{23}{3}}}$ (root of $x^3 - x - 1 = 0$). 
The bounds found led to Theorem~\ref{th:bound_dcycle_p} that concludes this part by highlighting that a vast majority of recurring configurations have the greatest minimum period possible.

\begin{theorem}[\cite{T-Noual2012}] Let $f$ be a \BAN{} of order $\omega$. 
  If $f$ is a \BAC{} or a \BADC{} such as $f$ is neither $\mathscr{D}_{5,1}^{-,-}$ nor $\mathscr{D}_{1,5}^{-,-}$, then its total number of attractors $\NBtotatt(\omega)$ is bounded by its total number of recurring configurations $\NBrec(\omega)$ so that:
  \begin{equation*} 
    \frac{\NBrec(\omega)}{\omega} \leq \NBtotatt(\omega) \leq 2 \cdot \frac{\NBrec(\omega)}{\omega} \text{,}
  \end{equation*}
  which means that the attractor periods of $f$ are very large:
  \begin{equation*}
    \sum_{p|\omega} p \cdot \frac{\NBatt(p)}{\NBtotatt(\omega)} = \frac{\NBrec(\omega)}{\NBtotatt(\omega)} \geq \frac{\omega}{2} \text{.}
  \end{equation*}
  \label{th:bound_dcycle_p}
\end{theorem}\smallskip

To end this section dedicated to parallel \BADCs{}, notice that all the results presented here extend naturally to any tangential double-cycles, namely two cycles that admit several automata in common such that these common automata are organised into an isolated path so that each of them have a local transition function of arity $1$ except the first one for which the arity is $2$. 
The proof is simple and rests on an induction consisting in transforming each tangential automaton whose local transition function arity is $1$ (by following the isolated path in reverse direction) into two copies, one in the left cycle and one in the right cycle.
Following this reasoning, it is easy to see that such a tangential double-cycle is equivalent to a \BADC{} whose cycles are of bigger sizes and same signs.\smallskip

Moreover, notice that both \BACs{} and \BADCs{} admit an exponential number of attractors. 
This is quite unrealistic if we view these objects as models of genetic regulation networks. 
Indeed, ``real'' genetic regulation networks seem to have a number of asymptotic behaviours (cellular types, biological rhythms...) that is polynomial (perhaps linear) according to the number of their genes. 
Nevertheless, it is important to see that the results obtained show that the number of attractors of \BADCs{} is drastically smaller than that of \BACs{}.
Actually, we think that the polynomial characteristics of the number of attractors of real regulation systems comes from the entanglement of cycles.  
More precisely, without formalising it, we conjecture that the more there are entangled cycles, the less there are local and global instabilities, the less there are attractors (simply because adding intersections adds dynamical constraints), and thus the less \ANs{} are sensitive to synchronism.

\subsection{Asynchronous \BADCs{}}
\label{sec:BADCs:a}

The last part of this synthesis is devoted to the dynamics of asynchronous \BADCs{}. It presents the  characterisation established in~\cite{C-Melliti2015}. 
It is
the counterpart of the study of the dynamics of \BADCs{} under  the parallel updating mode.
The study of the  asynchronous case takes a new approach. It formalises long sequences of updates by way of algorithmic descriptions. This approach  allows an elegant and more detailed description of the dynamics of feedbacks in interaction networks than the previous works. In particular it facilitates the study of convergence times.

\subsubsection{Definitions and notations}

For the sake of clarity, let us first recall that the study surveyed here focuses on canonical \BADCs{}.

\paragraph{\emph{States and configurations}}

Here, we will use the classical notation $V = \{0, \dots, n-1\}$ for representing the automata of a network of size $n$ and its congruence $V \equiv \{c = c_0, c_1, dots, c_{n-1}\}$. 
A configuration $x \in \BB^n$ is seen as a vector of two binary words. 
The first symbol of these two words represents $x_c \equiv x_0$. 
The null configuration is therefore denoted by $(\zero^\ell, \zero^r)$. Furthermore, we denote by $x^\ell$ (resp. $x^r$) the projection of $x$ on the \BAC{} $\mathscr{C}_\ell$ (resp. $\mathscr{C}_r$). 
This way, $x = (x^\ell, x^r)$, and in configuration $x$, the state of automaton $c_i^\ell$  is $x_i^\ell$. 
Notice that $x_0 = x_0^\ell = x_0^r$ since the three notations represent the state of automaton $c$ in $x$.

\paragraph{\emph{Expressiveness measure}}

Let $x$ be a configuration of a \BAC{} $\mathscr{C}_n$. 
Its \emph{expressiveness} is the number of factors $\zero\one$ that compose it, 
namely $|\{i\ |\ 0 \leq i \leq n - 1, x_i = \zero \text{ and } x_{i + 1 \mod n} = \one\}|$. 
The expressiveness of a configuration $x$ of a \BADC{} is the sum of the expressiveness of $x^\ell$ and $x^r$. 
From this definition follows that, if $\ell$ and $r$ are even, the least expressive configurations are $(\zero^\ell, \zero^r)$ and $(\one^\ell, \one^r)$, and that the most expressive ones are $((\zero\one)^\frac{\ell}{2}, (\zero\one)^\frac{r}{2})$ and $((\one\zero)^\frac{\ell}{2}, (\one\zero)^\frac{r}{2})$.

\paragraph{\emph{Elementary instructions}}

Network trajectories can be very long. 
To study them, we need a way to efficiently describe the sequence of automata updates that they execute. 
Our human minds must be able to understand from the description, what is the effect of the trajectory on the network, what changes does the network undergo along the trajectory. 
To do so, we proposed to view these sequences as instructions that make it easier to capture their effect on configurations.
Let us therefore consider: 
\begin{itemize}
\item a \BADC{} $\mathscr{D}_n$, 
\item one of the \BACs{} of $\mathscr{D}_n$, namely $\mathscr{C}$, whose size is noted  $\size {\mathscr{C}}$,
\item the current configuration $x$ of $\mathscr{C}$, and,
\item two automata of $\mathscr{C}$ distinct from $c$, namely $c_i$ and $c_j$,  such that $i < j$.
\end{itemize}
With these notations, the following seven basic instructions are defined:
\begin{enumerate}
\item \texttt{\small /* update of automaton $\com$ */}\\
  $\sync$: $x_c \leftarrow f_c(x)$
  \begin{itemize}
  \item[] Instruction $\sync$ is the only instruction that updates automaton $\com$   and where both \BACs{} interact with each other. 
    This (key)-instruction will always be called when $\com$ can change its state. 
    This instruction can be used either to set $\com$ at a desired state or to increase the expressiveness from a configuration. 
    Furthermore, it is the only way to switch a $\one\one\one$ (resp. $\zero\zero\zero$) pattern into a $\one\zero\one$ (resp. $\zero\one\zero$) pattern and, thus, to increase the expressiveness.
  \end{itemize}
\item \texttt{\small /* update of automaton $c_i$ */}\\
  $\update {c_i}$: $x_{c_i} \leftarrow f_{c_i}(x)$
  \begin{itemize}
  \item[] Instruction $\pupdate$ updates an automaton distinct from $\com$.
  \end{itemize}
\item \texttt{\small /* incremental updates */}\\
  $\clock {\mathscr{C}} {i} {j}$: \texttt{\small for} $k = i$ \texttt{\small upto} $j$ \texttt{\small do} $\update {c_k}$
  \begin{itemize}
  \item[] Instruction $\pclock$ updates consecutive automata in increasing order. 
    In fact, $\pclock$ propagates the state of $c_{i-1}$ along $\mathscr{C}$. Notice that if $j<i$ then no automata are updated. 
    Moreover, since $i \neq 0$ and $j \neq 0$, $\com$ cannot be updated with $\pclock$.
    This instruction admits the following property. 
    Let $x'$ be the result of the execution of $\clock {\mathscr{C}} {i} {j}$ on configuration $x$. 
    Then $\forall k \in \{i, \ldots, j\},\ x'_k = x_{i-1}$ and $\forall k \notin \{i,\ldots, j\},\ x'_k = x_k$.
  \end{itemize}
\item \texttt{\small /* incremental propagation of $x_c$ */}\\
  $\erase {\mathscr{C}}$: $\clock {\mathscr{C}} {1} {\size {\mathscr{C}} - 1}$
  \begin{itemize}
  \item[] Instruction $\perase$ is a particular case of $\pclock$.
    It propagates the state of $c_0$ along $\mathscr{C}$. 
    As a consequence, using $\perase$ on $\mathscr{C}$ decreases its expressiveness to $0$, and thus, is really efficient to reach quickly the least expressive configuration.
    This instruction admits the following property.
    Let $x'$ be the result of applying $\erase {\mathscr{C}}$ on configuration $x$.
    Then we have: $\forall k \in \{0, \ldots, \size {\mathscr{C}} -1\},\ x'_k  = x_{0}$.
  \end{itemize}
\item \texttt{\small /* incremental propagation of $x_c$ with no loss of expressiveness */}\\
  $\expand {\mathscr{C}}$: $\clock {\mathscr{C}} {1} {\kappa -1 \in \mathbb{N}}$\\ 
  where $\kappa = \underset{1 \leq k \leq \size {\mathscr{C} -1}}{\min} \left\lbrace k \ |\ \begin{cases}
    (x_k = 0) \text{ and } (x_{k+1 \mod \size {\mathscr{C}}} = 1) & \text{if } x_c = 1\\
    (x_k = 1) \text{ and } (x_{k+1 \mod \size {\mathscr{C}}} = 0) & \text{if } x_c = 0
  \end{cases} \right\rbrace$.
  \begin{itemize}
  \item[] Instruction $\pexpand$ is another particular case of $\pclock$ that aims   at propagating the state of $c_0$ along $\mathscr{C}$ while neither $\zero\one$   nor $\one\zero$ patterns are destroyed, which avoids decreasing the expressiveness of $\mathscr{C}$.
  \end{itemize}
\item \texttt{\small /* decremental updates */}\\
  $\counter {\mathscr{C}} {i} {j}$: \texttt{\small for} $k = j$ \texttt{\small downto} $i$ \texttt{\small do} $\update {c_k}$
  \begin{itemize}
  \item[] Instruction $\pcounter$ is the converse of instruction $\pclock$, and   updates consecutive automata in decreasing order. 
    Once $\counter {\mathscr{C}} {i} {j}$ executed, the information of $c_j$ is 
	lost and that of $c_{i-1}$ is possessed by both $c_{i-1}$ and $c_i$. 
	In fact, $\pcounter$ aims at shifting partially a \BAC{} section. 
	As for $\pclock$, if $j<i$ then no automata are updated and $\com$ cannot be updated with $\pcounter$.
    This instruction admits the following property. 
    Let $x'$ the result of the execution of $\counter {\mathscr{C}} {i} {j}$ on $x$. 
    Then $\forall k \in \{i, \ldots, j\},\ x'_k = x_{k-1}$ and $\forall k \notin \{i, \ldots, j\},\ x'_k = x_k$.
  \end{itemize}
\item \texttt{\small /* complete decremental update (except $c$) */}\\
  $\shift {\mathscr{C}}$: $\counter {\mathscr{C}} 1 {\size {\mathscr{C}} -1}$
  \begin{itemize}
  \item[] Instruction $\pshift$ is a particular case of instruction $\pcounter$. 
    Once executed, every automaton of $\mathscr{C}$ takes the state of its predecessor, except $\com$ whose state does not change. 
    Automaton $c_{\size {\mathscr{C} -1}}$ excluded, all the information contained along $\mathscr{C}$ is kept safe. 
    This instruction is useful to propagate information along a \BAC{} without loosing too much expressiveness (at most one $\zero\one$ pattern is destroyed).
  \end{itemize}
\end{enumerate}

\subsubsection{Results}

\paragraph{\emph{More complex instructions}}

Let $x$ be a configuration of a \BADC{} $\mathscr{D}$. 
Let us consider an algorithm composed of instructions defining an update sequence $\seq{x}$ from $x$. 
In every algorithm that follows, \BADC{} $\mathscr{D}$ is always considered as a global variable and is not mentioned.
Abusing language, $\seq{x}$ represents the sequence as well as the configuration 
resulting from its execution.

For the purpose of the study, we introduce three other more complex sequences in Table~\ref{tab:algo0}. 
In addition, Lemma~\ref{lem:copy} below shows  that the $\pcop$ instruction allows to transform $x$ into another configuration $x'$ if $x$ is sufficiently expressive.

\begin{table}[t!]
  \begin{center}
    \scalebox{1}{\centerline{
      \begin{minipage}{.46\textwidth}
        \centerline{
          \begin{tabular}{m{\textwidth}}
            \fbox{$\copc {x} {x'} {\mathscr{C}_m}$}\\[2mm]
            01.~ $\eta \leftarrow \size {\mathscr{C}_m}$;\\
            02.~ \textbf{\small if} ($x^m_{\eta-1} = x^m_{\eta-2} \textbf{ and } 
              x^m_{\eta-1} \neq x'^m_{\eta-1}$) \textbf{\small then}\\
            03.~ ~~~$j \leftarrow \max \{k\ |\ k < \eta-1 \text{ and } x^m_k \neq 
              x'^m_k\}$;\\
            04.~ \textbf{\small else}\\
            05.~ ~~~$j \leftarrow \eta$;\\
            06.~ \textbf{\small fi}\\
            07.~ \textbf{\small for} ($k = \eta-1$) 
              \textbf{\small downto} ($j + 1$) \textbf{\small do}\\
            08.~ ~~~$\update {c^m_{k-1}}$;\\
            09.~ ~~~$\update{c^m_{k}}$;\\
            10.~ \textbf{\small od}\\
            11.~ \textbf{\small for} ($k = j-1$) 
              \textbf{\small downto} ($1$) \textbf{\small do}\\
            12.~ ~~~\textbf{\small if} ($x^m_k \neq x'^m_k$) 
              \textbf{\small then}\\ 
            13.~ ~~~~~$\update{c^m_k}$;\\
            14.~ ~~~\textbf{\small fi}\\
            15.~ \textbf{\small od}
          \end{tabular}
        }
      \end{minipage}
      \hspace*{9mm}
      \begin{minipage}{.24\textwidth}
        \centerline{
          \begin{tabular}{m{\textwidth}}
            \fbox{$\cop {x} {x'}$}\\[2mm]
            01.~ $\copc {x} {x'} {\Cone}$;\\
            02.~ $\copc {x} {x'} {\Ctwo}$;
          \end{tabular}
        }
        \vspace*{5mm}
        \centerline{
          \begin{tabular}{m{\textwidth}}
            \fbox{$\copp {x} {x'}$}\\[2mm]
            01.~ \textbf{\small if} ($x_0 \neq x'_0$) \textbf{\small then}\\
            02.~ ~~~$\shift{\Cone}$;\\
            03.~ ~~~$\shift{\Ctwo}$;\\
            04.~ ~~~$\sync$;\\
            05.~ \textbf{\small fi}\\
            06.~ $\cop {x} {x'}$;
          \end{tabular}
        }
      \end{minipage}
    }}
  \caption{The update sequences $\pcopc$, $\pcop$ and $\pcopp$.}
  \label{tab:algo0}
  \end{center}
\end{table}

\begin{lemma}[\cite{C-Melliti2015}] Let $\mathscr{D}$ be a \BADC{} and let $x$ and $x'$ be two of its configurations such that $x_0 = x'_0$. If, for all $m \in \{\ell, r\}$, one of the following properties holds for $x$:
  \begin{enumerate}
  \item $\forall i \in \{1, \ldots, \size{\mathscr{C}_m} -1\},\ x^m_i \neq x^m_{i-1}$,
  \item $\forall i \in \{1, \ldots, \size{\mathscr{C}_m} -2\},\ x^m_i \neq x^m_{i-1} \text{ and } x^m_{\size{\mathscr{C}_m} -1} = x'^m_{\size{\mathscr{C}_m} -1}$,
  \item $\forall i \in \{1, \ldots, \size{\mathscr{C}_m} -2\},\ x^m_i \neq x^m_{i-1} \text{ and } \exists p \in \{1, \ldots, \size{\mathscr{C}_m} -2\}, x^m_p \neq x'^m_p$,
  \end{enumerate}
  then $\cop {x} {x'} = x'$ and this sequence executes at most $2(\ell + r - 6)$ updates.
  \label{lem:copy}
\end{lemma}

This lemma gives strong insights about the expressive power of instructions and sequences to reveal possible trajectories between configurations. 
Now let us focus on the dynamical behaviour of \BADCs{}, from a general point of view.

\paragraph{\emph{Positive \BADCs{}}}

In section~\ref{sec:BADCs:p}, we have seen that positive \BADCs{} behave like positive \BACs{}, namely they have two stable configurations among their 
attractors. 
In the asynchronous case, these two stable configurations are the only attractors. 
The general idea of the demonstration is based on canonical positive \BADCs{} and establishes that the two sequences $\pconvn$ and $\pconvp$ given in Table~\ref{tab:algo1} allow to transform any configuration with at least one automaton in state $\zero$ into configuration $(\zero^\ell, \zero^r)$, and any configuration with at least one automaton in state $\one$ in each of its \BACs{} into configuration $(\one^\ell, \one^r)$. 
This result is illustrated at the top of Figure~\ref{fig:adcycles_tg}, and is  summarised in Theorem~\ref{th:pbadc} below.

\begin{table}[t!]
  \begin{center}
    \scalebox{1}{\centerline{
      \begin{minipage}{.31\textwidth}
        \centerline{
          \begin{tabular}{m{\textwidth}}
            \fbox{$\convn {x}$}\\[2mm]
            01.~ \textbf{\small if} ($x_0 = \one$) \textbf{\small then}\\
            02.~ ~~~$i \leftarrow \min \{k\ |\ x^\ell_k = \zero\}$;\\
            03.~ ~~~$\clock {\Cone} {i+1} {\ell-1}$;\\
            04.~ ~~~$\sync$;\\
            05.~ \textbf{\small fi}\\
            06.~ $\erase {\Cone}$;\\
            07.~ $\erase {\Ctwo}$;
          \end{tabular}
        }
      \end{minipage}
      \hspace*{10mm}
      \begin{minipage}{.32\textwidth}
        \centerline{
          \begin{tabular}{m{\textwidth}}
            \fbox{$\convp {x}$}\\[2mm]
            01.~ \textbf{\small if} ($x_0 = \zero$) \textbf{\small then}\\
            02.~ ~~~$i \leftarrow \min \{k\ |\ x^\ell_k = \one\}$;\\
            03.~ ~~~$\clock {\Cone} {i+1} {\ell-1}$;\\
            04.~ ~~~$j \leftarrow \min \{k\ |\ x^r_k = \one\}$;\\
            05.~ ~~~$\clock {\Ctwo} {j+1} {r-1}$;\\
            06.~ ~~~$\sync$;\\
            07.~ \textbf{\small fi}\\
            08.~ $\erase {\Cone}$;\\
            09.~ $\erase {\Ctwo}$;
          \end{tabular}
        }
      \end{minipage}
	}}
    \caption{The update sequences $\pconvn$ and $\pconvp$.}
    \label{tab:algo1}
  \end{center}
\end{table}

\begin{theorem}[\cite{C-Melliti2015}] Let $\mathscr{D}^{+,+}$ be a canonical positive \BADC{} where $\diamond = \land$ and $x$ one of its unstable configurations. 
  If $x$ admits one automaton in state $0$, then $\convn {x} = (\zero^\ell, \zero^r)$. 
  Moreover, if in configuration $x$, there is  one automaton in state $\one$ in each \BAC{}, then $\convp {x} = (\one^\ell, \one^r)$. 
  The convergence time of $\mathscr{D}^{+,+}$ is at most $2(\ell + r) - 5$.
  \label{th:pbadc}
\end{theorem}

\paragraph{\emph{Mixed \BADCs{}}}

For \textit{mixed} asynchronous \BADCs{}, as for positive ones, asynchronism allows to eliminate all local instabilities. 
Thus, contrary to parallel \BADCs{}, asynchronous \BADCs{} have only one attractor that is a stable configuration. 
In the canonical case, this is evidenced by the $\psimp$ sequence given in Table~\ref{tab:algo2} that provides a way to converge towards this stable configuration from any initial configuration $x$, by reducing progressively its expressiveness. 
This result is illustrated in Figure~\ref{fig:adcycles_tg} (middle) and is formalised by Theorem~\ref{th:mbadc} below.

\begin{theorem}[\cite{C-Melliti2015}] Let $\mathscr{D}^{-,+}$ be a canonical mixed  \BADC{}, where $\diamond = \land$. 
  For any of its configuration $x$, $\simp {x} = (\zero^\ell,\zero^r)$ holds. 
  The convergence time of $\mathscr{D}^{-,+}$ is at most $2\ell + r - 2$.  
  \label{th:mbadc}
\end{theorem}

\paragraph{\emph{Negative \BADCs{}}}

Here, we distinguish between (1) \emph{even} negative \BADCs{} and (2) \emph{odd} negative \BADCs{}. 
The first ones are defined as having two even-sized \BACs{}, the second ones as having at least one odd-sized \BAC{}.\smallskip

Even negative \BADCs{} admit a single attractor. 
This attractor is a stable oscillation of length $2^{\ell + r - 1}$. 
This means that all the configurations are recurring and consequently the convergence time is null. 
All configurations are reachable. 
However this result also means that configurations of maximal expressiveness are hard to reach: the number of updates to reach them is quadratic according to the size of the \BADC{}.
The general idea of the proof follows the following three points:
\begin{itemize}
\item Any configuration can reach the less expressive one $\allzero$ in linear time. \hfill (P1)
\item Configuration $\allzero $ can reach the highest expressive one $\altone$ in quadratic time. \hfill (P2)
\item Any configuration can be reached from $\altone$ in linear time. \hfill (P3)
\end{itemize}\smallskip

\begin{table}[t!]
  \begin{center}
    \scalebox{1}{\centerline{
      \begin{minipage}{.2\textwidth}
        \centerline{
          \begin{tabular}{m{\textwidth}}
            \fbox{$\simp {x}$}\\[2mm]
            01.~ \textbf{\small if} ($x_0 = \one$) \textbf{\small then}\\
            02.~ ~~~$\erase {\Cone}$;\\
            03.~ ~~~$\sync$;\\
            04.~ \textbf{\small fi}\\
            05.~ $\erase {\Cone}$;\\
            06.~ $\erase {\Ctwo}$;
          \end{tabular}
        }
      \end{minipage}
      \hspace*{7.5mm}
      \begin{minipage}{.31\textwidth}
        \centerline{
          \begin{tabular}{m{\textwidth}}
            \fbox{$\compa {x}$}\\[2mm]
            01.~ \textbf{\small for} ($i = 1$) 
              \textbf{\small upto} ($\ell-1$) \textbf{\small do}\\
            02.~ ~~~ $\sync$;\\
            03.~ ~~~ $\expand {\Cone}$;\\
            04.~ ~~~ $\erase {\Ctwo}$;\\
            05.~ \textbf{\small od}
          \end{tabular}
        }
      \end{minipage}
      \hspace*{7.5mm}
      \begin{minipage}{.31\textwidth}
        \centerline{
          \begin{tabular}{m{\textwidth}}
            \fbox{$\compb {x}$}\\[2mm]
            01.~ \textbf{\small if} ($x^r= \one^r$) \textbf{\small then}\\
            02.~ ~~~$\sync$;\\
            03.~ ~~~$\erase {\Ctwo}$;\\
            04.~ \textbf{\small fi}\\
            05.~ $\sync$;\\
            06.~ $\expand {\Ctwo}$;\\
            07.~ \textbf{\small for} ($i = 1$) \textbf{\small upto} ($r-2$) 
              \textbf{\small do}\\
            08.~ ~~~$\shift {\Cone}$;\\
            09.~ ~~~$\sync$;\\
            10.~ ~~~$\expand {\Ctwo}$;\\
            11.~ \textbf{\small od}
          \end{tabular}
        }
      \end{minipage}
	}}
	\caption{The update sequences $\psimp$, $\pcompa$ and $\pcompb$.}
	\label{tab:algo2}
  \end{center}
\end{table}

Consider P1. It is easy to see that the sequence $\psimp$ remains effective for
reaching $\allzero$, which is formalised by Lemma~\ref{lem:simply} below.

\begin{lemma}[\cite{C-Melliti2015}] For any configuration $x$ of $\mathscr{D}_{\ell, r}^{-,-}$, $\simp{x} = \allzero$ and executes at most $2\ell + r - 2 $ updates.
  \label{lem:simply}
\end{lemma}

\begin{figure}[t!]
  \begin{center}
    \begin{minipage}{.45\textwidth}
      \centerline{\scalebox{1}{\begin{tikzpicture}[>=to,auto]
        \tikzstyle{type} = []
        \tikzstyle{conf} = [rectangle, draw]
        \tikzstyle{pf} = [rectangle, thick, draw, fill=black!20]
        \tikzstyle{lc} = [rectangle, thick, draw, fill=black!80]
        \node[type] (d) at (0,3.75) {$\mathscr{D}_{2,2}^{+,+}$};
        \node[pf](n000) at (0,0) {$\zero\zero\zero$};
        \node[conf](n001) at (2.5,0) {$\zero\zero\one$};
        \node[conf](n010) at (0,2.5) {$\zero\one\zero$};
        \node[conf](n011) at (2.5,2.5) {$\zero\one\one$};
        \node[conf](n100) at (1.25,1.25) {$\one\zero\zero$};
        \node[conf](n101) at (3.75,1.25) {$\one\zero\one$};
        \node[conf](n110) at (1.25,3.75) {$\one\one\zero$};
        \node[pf](n111) at (3.75,3.75) {$\one\one\one$};
        \draw[thick,Green4,->] (n000) edge [loop left, distance=5mm] (n000);
        \draw[thick,Blue3,->] (n000) edge [loop left, distance=7mm] (n000);
        \draw[thick,Red3,->] (n000) edge [loop left, distance=9mm] (n000);
        \draw[thick,Green4,->] (n001) edge (n000);
        \draw[thick,Blue3,->] (n001) edge [loop right, distance=5mm] (n001);
        \draw[thick,Red3,->] (n001) edge [loop right, distance=7mm] (n001);
        \draw[thick,Green4,->] (n010) edge (n011);
        \draw[thick,Blue3,->] (n010) edge (n000);
        \draw[thick,Red3,->] (n010) edge (n110);
        \draw[thick,Green4,->] (n011) edge [loop right, distance=5mm] (n011);
        \draw[thick,Blue3,->] (n011) edge  (n001);
        \draw[thick,Red3,->] (n011) edge (n111);
        \draw[thick,Green4,->] (n100) edge [loop left, distance=5mm] (n100);
        \draw[thick,Blue3,->] (n100) edge [loop left, distance=7mm] (n100);
        \draw[thick,Red3,->] (n100) edge (n000);        
        \draw[thick,Green4,->] (n101) edge (n100);
        \draw[thick,Blue3,->] (n101) edge (n111);
        \draw[thick,Red3,->] (n101) edge (n001);
        \draw[thick,Green4,->] (n110) edge (n111);
        \draw[thick,Blue3,->] (n110) edge (n100);
        \draw[thick,Red3,->] (n110) edge [loop left, distance=5mm] (n110);
        \draw[thick,Green4,->] (n111) edge [loop right, distance=5mm] (n111);
        \draw[thick,Blue3,->] (n111) edge [loop right, distance=7mm] (n111);
        \draw[thick,Red3,->] (n111) edge [loop right, distance=9mm] (n111);
      \end{tikzpicture}}}
    \end{minipage}
    \quad\vrule\quad
    \begin{minipage}{.45\textwidth}
      \centerline{\scalebox{1}{\begin{tikzpicture}[>=to,auto]
        \tikzstyle{type} = []
        \tikzstyle{conf} = [rectangle, draw]
        \tikzstyle{pf} = [rectangle, thick, draw, fill=black!20]
        \tikzstyle{lc} = [rectangle, thick, draw, fill=black!80]
        \node[type] (d) at (0,3.75) {$\mathscr{D}_{1,3}^{+,+}$};
        \node[pf](n000) at (0,0) {$\zero\zero\zero$};
        \node[conf](n001) at (2.5,0) {$\zero\zero\one$};
        \node[conf](n010) at (0,2.5) {$\zero\one\zero$};
        \node[conf](n011) at (2.5,2.5) {$\zero\one\one$};
        \node[conf](n100) at (1.25,1.25) {$\one\zero\zero$};
        \node[conf](n101) at (3.75,1.25) {$\one\zero\one$};
        \node[conf](n110) at (1.25,3.75) {$\one\one\zero$};
        \node[pf](n111) at (3.75,3.75) {$\one\one\one$};
        \draw[thick,Green4,->] (n000) edge [loop left, distance=5mm] (n000);
        \draw[thick,Blue3,->] (n000) edge [loop left, distance=7mm] (n000);
        \draw[thick,Red3,->] (n000) edge [loop left, distance=9mm] (n000);
        \draw[thick,Green4,->] (n001) edge (n000);
        \draw[thick,Blue3,->] (n001) edge [loop right, distance=5mm] (n001);
        \draw[thick,Red3,->] (n001) edge [loop right, distance=7mm] (n001);
        \draw[thick,Green4,->] (n010) edge (n011);
        \draw[thick,Blue3,->] (n010) edge (n000);
        \draw[thick,Red3,->] (n010) edge [loop left, distance=5mm] (n010);
        \draw[thick,Green4,->] (n011) edge [loop right, distance=5mm] (n011);
        \draw[thick,Blue3,->] (n011) edge (n001);
        \draw[thick,Red3,->] (n011) edge [loop right, distance=7mm] (n011);
        \draw[thick,Green4,->] (n100) edge [loop left, distance=5mm] (n100);
        \draw[thick,Blue3,->] (n100) edge (n110);
        \draw[thick,Red3,->] (n100) edge (n000);        
        \draw[thick,Green4,->] (n101) edge (n100);
        \draw[thick,Blue3,->] (n101) edge (n111);
        \draw[thick,Red3,->] (n101) edge [loop right, distance=5mm] (n101);
        \draw[thick,Green4,->] (n110) edge (n111);
        \draw[thick,Blue3,->] (n110) edge [loop left, distance=5mm] (n110);
        \draw[thick,Red3,->] (n110) edge (n010);
        \draw[thick,Green4,->] (n111) edge [loop right, distance=5mm] (n111);
        \draw[thick,Blue3,->] (n111) edge [loop right, distance=7mm] (n111);
        \draw[thick,Red3,->] (n111) edge [loop right, distance=9mm] (n111);
      \end{tikzpicture}}}
    \end{minipage}\bigskip
    
    \hrule\bigskip
    
    \begin{minipage}{.45\textwidth}
      \centerline{\scalebox{1}{\begin{tikzpicture}[>=to,auto]
        \tikzstyle{type} = []
        \tikzstyle{conf} = [rectangle, draw]
        \tikzstyle{pf} = [rectangle, thick, draw, fill=black!20]
        \tikzstyle{lc} = [rectangle, thick, draw, fill=black!80]
        \node[type] (d) at (0,3.75) {$\mathscr{D}_{2,2}^{-,+}$};
        \node[pf](n000) at (0,0) {$\zero\zero\zero$};
        \node[conf](n001) at (2.5,0) {$\zero\zero\one$};
        \node[conf](n010) at (0,2.5) {$\zero\one\zero$};
        \node[conf](n011) at (2.5,2.5) {$\zero\one\one$};
        \node[conf](n100) at (1.25,1.25) {$\one\zero\zero$};
        \node[conf](n101) at (3.75,1.25) {$\one\zero\one$};
        \node[conf](n110) at (1.25,3.75) {$\one\one\zero$};
        \node[conf](n111) at (3.75,3.75) {$\one\one\one$};
        \draw[thick,Green4,->] (n000) edge [loop left, distance=5mm] (n000);
        \draw[thick,Blue3,->] (n000) edge [loop left, distance=7mm] (n000);
        \draw[thick,Red3,->] (n000) edge [loop left, distance=9mm] (n000);
        \draw[thick,Green4,->] (n001) edge (n000);
        \draw[thick,Blue3,->] (n001) edge (n011);
        \draw[thick,Red3,->] (n001) edge [loop right, distance=5mm] (n001);
        \draw[thick,Green4,->] (n010) edge (n011);
        \draw[thick,Blue3,->] (n010) edge (n000);
        \draw[thick,Red3,->] (n010) edge (n110);
        \draw[thick,Green4,->] (n011) edge [loop right, distance=5mm] (n011);
        \draw[thick,Blue3,->] (n011) edge [loop right, distance=7mm] (n011);
        \draw[thick,Red3,->] (n011) edge (n111);
        \draw[thick,Green4,->] (n100) edge [loop left, distance=5mm] (n100);
        \draw[thick,Blue3,->] (n100) edge [loop left, distance=7mm] (n100);
        \draw[thick,Red3,->] (n100) edge (n000);        
        \draw[thick,Green4,->] (n101) edge (n100);
        \draw[thick,Blue3,->] (n101) edge [loop right, distance=5mm] (n101);
        \draw[thick,Red3,->] (n101) edge (n001);
        \draw[thick,Green4,->] (n110) edge (n111);
        \draw[thick,Blue3,->] (n110) edge (n100);
        \draw[thick,Red3,->] (n110) edge [loop left, distance=5mm] (n110);
        \draw[thick,Green4,->] (n111) edge [loop right, distance=5mm] (n111);
        \draw[thick,Blue3,->] (n111) edge (n101);
        \draw[thick,Red3,->] (n111) edge [loop right, distance=7mm] (n111);
      \end{tikzpicture}}}
    \end{minipage}
    \quad\vrule\quad
    \begin{minipage}{.45\textwidth}
      \centerline{\scalebox{1}{\begin{tikzpicture}[>=to,auto]
        \tikzstyle{type} = []
        \tikzstyle{conf} = [rectangle, draw]
        \tikzstyle{pf} = [rectangle, thick, draw, fill=black!20]
        \tikzstyle{lc} = [rectangle, thick, draw, fill=black!80]
        \node[type] (d) at (0,3.75) {$\mathscr{D}_{1,3}^{-,+}$};
        \node[pf](n000) at (0,0) {$\zero\zero\zero$};
        \node[conf](n001) at (2.5,0) {$\zero\zero\one$};
        \node[conf](n010) at (0,2.5) {$\zero\one\zero$};
        \node[conf](n011) at (2.5,2.5) {$\zero\one\one$};
        \node[conf](n100) at (1.25,1.25) {$\one\zero\zero$};
        \node[conf](n101) at (3.75,1.25) {$\one\zero\one$};
        \node[conf](n110) at (1.25,3.75) {$\one\one\zero$};
        \node[conf](n111) at (3.75,3.75) {$\one\one\one$};
        \draw[thick,Green4,->] (n000) edge [loop left, distance=5mm] (n000);
        \draw[thick,Blue3,->] (n000) edge [loop left, distance=7mm] (n000);
        \draw[thick,Red3,->] (n000) edge [loop left, distance=9mm] (n000);
        \draw[thick,Green4,->] (n001) edge (n000);
        \draw[thick,Blue3,->] (n001) edge [loop right, distance=5mm] (n001);
        \draw[thick,Red3,->] (n001) edge [bend left=10] (n101);
        \draw[thick,Green4,->] (n010) edge (n011);
        \draw[thick,Blue3,->] (n010) edge (n000);
        \draw[thick,Red3,->] (n010) edge [loop left, distance=5mm] (n010);
        \draw[thick,Green4,->] (n011) edge [loop right, distance=5mm] (n011);
        \draw[thick,Blue3,->] (n011) edge (n001);
        \draw[thick,Red3,->] (n011) edge [bend left=10] (n111);
        \draw[thick,Green4,->] (n100) edge [loop left, distance=5mm] (n100);
        \draw[thick,Blue3,->] (n100) edge (n110);
        \draw[thick,Red3,->] (n100) edge (n000);        
        \draw[thick,Green4,->] (n101) edge (n100);
        \draw[thick,Blue3,->] (n101) edge (n111);
        \draw[thick,Red3,->] (n101) edge [bend left=10] (n001);
        \draw[thick,Green4,->] (n110) edge (n111);
        \draw[thick,Blue3,->] (n110) edge [loop left, distance=5mm] (n110);
        \draw[thick,Red3,->] (n110) edge (n010);
        \draw[thick,Green4,->] (n111) edge [loop right, distance=5mm] (n111);
        \draw[thick,Blue3,->] (n111) edge [loop right, distance=7mm] (n111);
        \draw[thick,Red3,->] (n111) edge [bend left=10] (n011);      
        \end{tikzpicture}}}
    \end{minipage}\bigskip

    \hrule\bigskip
    
    \begin{minipage}{.45\textwidth}
      \centerline{\scalebox{1}{\begin{tikzpicture}[>=to,auto]
        \tikzstyle{type} = []
        \tikzstyle{conf} = [rectangle, draw]
        \tikzstyle{pf} = [rectangle, thick, draw, fill=black!20]
        \tikzstyle{lc} = [rectangle, thick, draw, fill=black!80]
        \node[type] (d) at (0,3.75) {$\mathscr{D}_{2,2}^{-,-}$};
        \node[lc](n000) at (0,0) {\textcolor{white}{$\zero\zero\zero$}};
        \node[lc](n001) at (2.5,0) {\textcolor{white}{$\zero\zero\one$}};
        \node[lc](n010) at (0,2.5) {\textcolor{white}{$\zero\one\zero$}};
        \node[lc](n011) at (2.5,2.5) {\textcolor{white}{$\zero\one\one$}};
        \node[lc](n100) at (1.25,1.25) {\textcolor{white}{$\one\zero\zero$}};
        \node[lc](n101) at (3.75,1.25) {\textcolor{white}{$\one\zero\one$}};
        \node[lc](n110) at (1.25,3.75) {\textcolor{white}{$\one\one\zero$}};
        \node[lc](n111) at (3.75,3.75) {\textcolor{white}{$\one\one\one$}};
        \draw[thick,Green4,->] (n000) edge [loop left, distance=5mm] (n000);
        \draw[thick,Blue3,->] (n000) edge (n010);
        \draw[thick,Red3,->] (n000) edge [loop left, distance=7mm] (n000);
        \draw[thick,Green4,->] (n001) edge (n000);
        \draw[thick,Blue3,->] (n001) edge [loop right, distance=5mm] (n001);
        \draw[thick,Red3,->] (n001) edge [loop right, distance=7mm] (n001);
        \draw[thick,Green4,->] (n010) edge (n011);
        \draw[thick,Blue3,->] (n010) edge [loop left, distance=5mm] (n010);
        \draw[thick,Red3,->] (n010) edge (n110);
        \draw[thick,Green4,->] (n011) edge [loop right, distance=5mm] (n011);
        \draw[thick,Blue3,->] (n011) edge (n001);
        \draw[thick,Red3,->] (n011) edge (n111);
        \draw[thick,Green4,->] (n100) edge [loop left, distance=5mm] (n100);
        \draw[thick,Blue3,->] (n100) edge [loop left, distance=7mm] (n100);
        \draw[thick,Red3,->] (n100) edge (n000);
        \draw[thick,Green4,->] (n101) edge (n100);
        \draw[thick,Blue3,->] (n101) edge [loop right, distance=5mm] (n101);
        \draw[thick,Red3,->] (n101) edge (n001);
        \draw[thick,Green4,->] (n110) edge (n111);
        \draw[thick,Blue3,->] (n110) edge (n100);
        \draw[thick,Red3,->] (n110) edge [loop left, distance=5mm] (n110);
        \draw[thick,Green4,->] (n111) edge [loop right, distance=5mm] (n111);
        \draw[thick,Blue3,->] (n111) edge (n101);
        \draw[thick,Red3,->] (n111) edge [loop right, distance=7mm] (n111);      
      \end{tikzpicture}}}
    \end{minipage}
    \quad\vrule\quad
    \begin{minipage}{.45\textwidth}
      \centerline{\scalebox{1}{\begin{tikzpicture}[>=to,auto]
        \tikzstyle{type} = []
        \tikzstyle{conf} = [rectangle, draw]
        \tikzstyle{pf} = [rectangle, thick, draw, fill=black!20]
        \tikzstyle{lc} = [rectangle, thick, draw, fill=black!80]
        \node[type] (d) at (0,3.75) {$\mathscr{D}_{1,3}^{-,-}$};
        \node[lc](n000) at (0,0) {\textcolor{white}{$\zero\zero\zero$}};
        \node[lc](n001) at (2.5,0) {\textcolor{white}{$\zero\zero\one$}};
        \node[lc](n010) at (0,2.5) {\textcolor{white}{$\zero\one\zero$}};
        \node[lc](n011) at (2.5,2.5) {\textcolor{white}{$\zero\one\one$}};
        \node[lc](n100) at (1.25,1.25) {\textcolor{white}{$\one\zero\zero$}};
        \node[conf](n101) at (3.75,1.25) {$\one\zero\one$};
        \node[lc](n110) at (1.25,3.75) {\textcolor{white}{$\one\one\zero$}};
        \node[lc](n111) at (3.75,3.75) {\textcolor{white}{$\one\one\one$}};
        \draw[thick,Green4,->] (n000) edge [loop left, distance=5mm] (n000);
        \draw[thick,Blue3,->] (n000) edge [loop left, distance=7mm] (n000);
        \draw[thick,Red3,->] (n000) edge [bend left=10] (n100);
        \draw[thick,Green4,->] (n001) edge (n000);
        \draw[thick,Blue3,->] (n001) edge [loop right, distance=5mm] (n001);
        \draw[thick,Red3,->] (n001) edge [loop right, distance=7mm] (n001);
        \draw[thick,Green4,->] (n010) edge (n011);
        \draw[thick,Blue3,->] (n010) edge (n000);
        \draw[thick,Red3,->] (n010) edge [bend left=10](n110);
        \draw[thick,Green4,->] (n011) edge [loop right, distance=5mm] (n011);
        \draw[thick,Blue3,->] (n011) edge (n001);
        \draw[thick,Red3,->] (n011) edge [loop right, distance=7mm] (n011);
        \draw[thick,Green4,->] (n100) edge [loop left, distance=5mm] (n100);
        \draw[thick,Blue3,->] (n100) edge (n110);
        \draw[thick,Red3,->] (n100) edge [bend left=10] (n000);        
        \draw[thick,Green4,->] (n101) edge (n100);
        \draw[thick,Blue3,->] (n101) edge (n111);
        \draw[thick,Red3,->] (n101) edge (n001);
        \draw[thick,Green4,->] (n110) edge (n111);
        \draw[thick,Blue3,->] (n110) edge [loop left, distance=5mm] (n110);
        \draw[thick,Red3,->] (n110) edge [bend left=10] (n010);
        \draw[thick,Green4,->] (n111) edge [loop right, distance=5mm] (n111);
        \draw[thick,Blue3,->] (n111) edge [loop right, distance=7mm] (n111);
        \draw[thick,Red3,->] (n111) edge (n011);      
      \end{tikzpicture}}}
    \end{minipage}\smallskip
  \caption{Asynchronous transition graphs of the canonical \BADCs{} depicted in Figure~\ref{fig:pdcycles_ig}, where $\diamond = \land$ and where every $\textcolor{Red3}{\to}$ (resp. $\textcolor{Blue3}{\to}$ and $\textcolor{Green4}{\to}$) represents an update of automaton $0$ (resp. $1$ and $2$).}
  \label{fig:adcycles_tg}
  \end{center}
\end{figure}
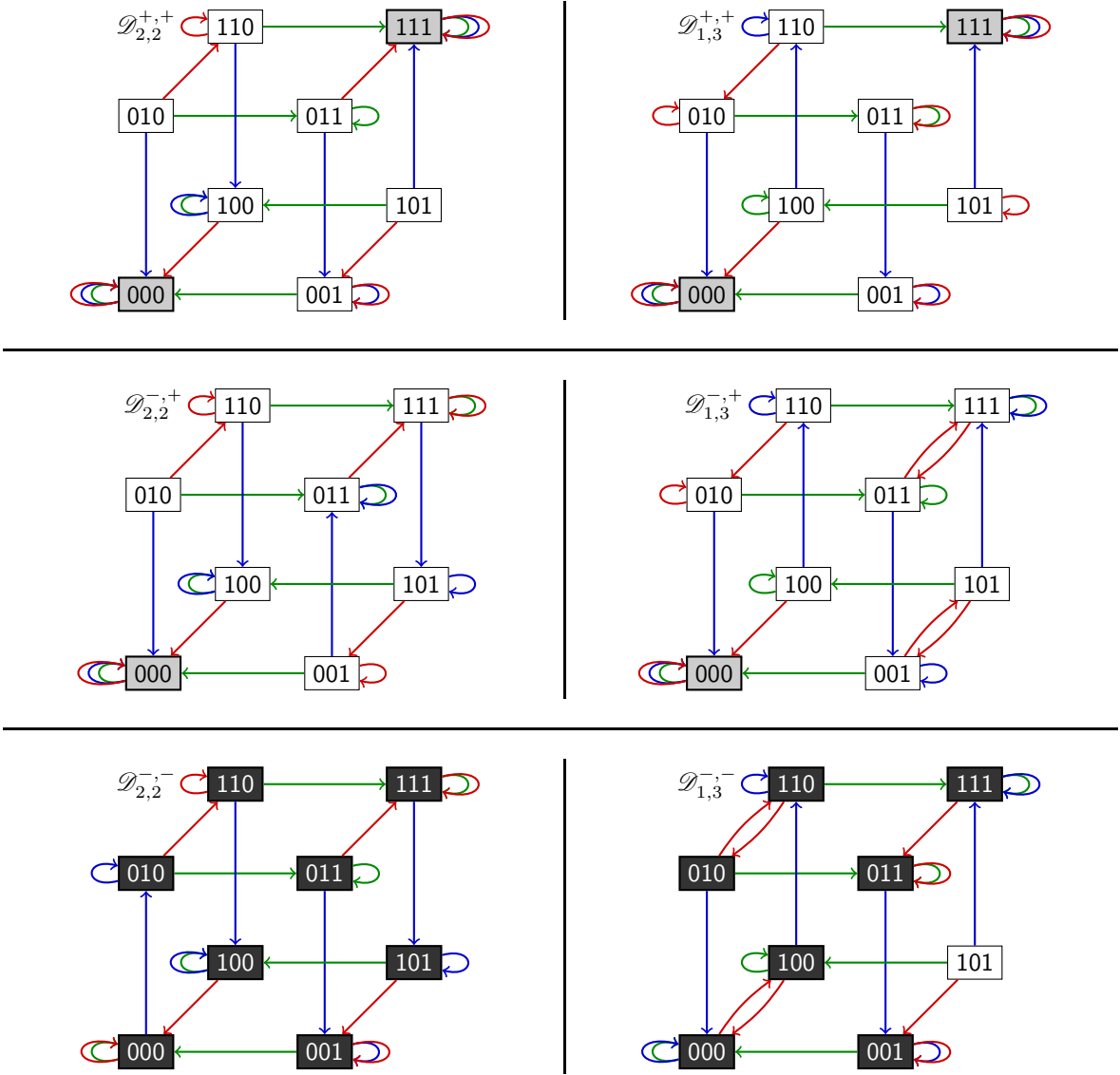

Now, consider P2. 
Implicitly, P2 requires increasing the expressiveness of $\allzero$ by successive updates, and finding a trajectory that reaches $\altone$. 
To do so, we proceed in two stages. 
First, we increase the expressiveness of $\Cone$ using $\pcompa$ (see Lemma~\ref{lem:complex1}). 
Then, we increase the expressiveness of $\Ctwo$.  
This second stage is carried out without reducing the expressiveness of $\Cone$ by using $\pcompb$ (see Lemma~\ref{lem:complex2}). 
The expected result follows from the composition $\pcomp = \pcompb \circ \pcompa$.
It is formalised in Lemma~\ref{lem:complex}.

\begin{lemma}[\cite{C-Melliti2015}] In an even negative \BADC{} $\mathscr{D}_{\ell,r}^{-,-}$, sequence $$\compa {\allzero}$$ leads to configuration $((\one \zero)^{\frac{\ell}{2}}, \one^r)$ and executes at most $ (\ell-1) (\ell + r - 2)$ updates. 
  \label{lem:complex1}
\end{lemma}

\begin{lemma}[\cite{C-Melliti2015}] In an even negative \BADC{} $\mathscr{D}_{\ell,r}^{-,-} $, sequence $$\compb {((\one\zero)^{\frac{\ell}{2}}, \one^r)}$$ leads to configuration $\altone$ and executes at most $(r-2) (\ell + r - 2) + (2r-1)$ updates.
  \label{lem:complex2}
\end{lemma}

\begin{lemma}[\cite{C-Melliti2015}] In an even negative \BADC{} $\mathscr{D}_{\ell,r}^{-,-}$, sequence $$\comp {\allzero}$$ leads to configuration $\altone$ and executes at most $ (\ell + r) ^ 2 - 5 (\ell - 1) - 3r$ updates.
  \label{lem:complex}
\end{lemma}

P3 is developed in Lemma~\ref{lem:copyPair} that uses the $\pcopp$ sequence (see 
Table~\ref{tab:algo0}).

\begin{lemma}[\cite{C-Melliti2015}] In an even negative \BADC{} $\mathscr{D}_{\ell,r}^{-,-}$, for any configuration $x'$, sequence $$\copp {\altone} {x'}$$ transforms configuration $\altone$ into $x'$ in at most $3 (\ell + r - 4) - 1$ updates.
  \label{lem:copyPair}
\end{lemma}

Starting from Lemmas~\ref{lem:simply} to~\ref{lem:copyPair}, whatever the  configurations $x$ and $x'$, the composition $$\copp {\comp {\simp {x}}} {x'} = x'$$ holds, which proves that there is a unique attractor of length $2^{\ell+r-1}$. 
From this is derived Theorem~\ref{thm:oneAttEven}. It gives some bounds on convergence time.

\begin{theorem}[\cite{C-Melliti2015}] Let $\mathscr{D}_{\ell,r}^{-,-}$ be a canonical negative \BADC{}, where $\diamond = \land$. $\mathscr{D}_{\ell,r}^{-,-}$ admits a unique attractor of length $2^{\ell+r-1}$. 
  In this stable oscillation, any configuration can be reached from any other in $O(\ell^2 + r^2)$ updates. 
  However, configurations $\allzero$ and $\allone$ can be reached from any other one in $O(\ell+r)$ updates, and the configurations $\altzero$ and $\altone$ can reach all the others in $O(\ell+r)$ updates.
  \label{thm:oneAttEven}
\end{theorem}

Like even negative \BADCs{}, odd negative \BADCs{} also admit a single attractor (a stable oscillation). However, all configurations are not necessarily recurring. 
Indeed, they admit a set $I$ of non-reachable configurations, from which updates are irreversible. 
The associated result is formalised in Theorem~\ref{thm:sizeof} below.

\begin{theorem}[\cite{C-Melliti2015}] Let $\rho: \mathbb {N} \to \{0,1\} $, with $\rho(k) = \begin{cases}
    0 & \text{ if } k = 0 \text{ or } k \equiv 1 \mod 2 \\
    1 & \text{otherwise}
  \end{cases}$. 
  Every \BADC{} $\mathscr{D}_{\ell, r}^{-, -} $ admits a unique attractor $\BB^{\ell + r-1} \setminus I$, where $|I| = \alpha(\ell - 1) \times 2^{r - 1} + \alpha(r - 1) \times 2^{\ell - 1}$.
  \label{thm:sizeof}
\end{theorem}

This result, whose proof rests on the characterisation of $I$, generalises Theorem~\ref{thm:oneAttEven}.

\section{Conclusion}
\label{sec:conclusion}

In this chapter, we have summarised the major results obtained in recent years on the role of feedbacks in interaction networks. 
The literature has already established that feedback patterns are "engines of complexity" in the dynamical behaviours of larger networks that contain them.
Because of their proven essential character, we have therefore deliberately focused on the dynamics of cycles and double-cycles, by focusing especially on the influence of updating modes. 
Without going back on the results themselves, this chapter highlights the fundamental differences induced by the ``scheduling'' of events on the behaviours of the complexity engines of interaction networks: 
while parallelism tends to render all the asymptotic mathematical diversity of 
networks, pure asynchronism tends to reduce degrees of freedom of networks, which 
partly explains why the former is often preferred in theoretical work while the 
second is often adopted in works that are oriented towards applications in 
molecular biology. 
At present, no real knowledge in molecular biology establishes precisely how regulations are implemented over time, even if works state that the chromatin dynamics plays an important role. 
That is in particular why the type of studies developed in this chapter, focusing on updating modes, remains essential to go further in understanding formal and applied interaction networks. 
As a consequence, further studies need to be done on the scheduling of updates over time.
A first avenue is to focus for instance on more likely updating modes, in agreement with the discussion of what is claimed in~\cite{J-Demongeot2020}.\medskip

Another natural opening highlighted by these works, deeply rooted in fundamental computer science, consists in developing knowledge on the dynamical properties of interaction networks, including relationships between their architecture and structure. 
But other equally relevant lines of investigation also call for exploration.
One of them was opened up by the study of asynchronous double-cycles and deals with the time complexity of networks. 
A question that remains currently open is the following: does the time complexity of the networks go hand in hand with their behavioural diversity? 
If yes, can we find a measure of it? 
Finally, the work carried out on cycles and double-cycles in parallel, developed in ~\cite{T-Noual2012, T-Sene2012} discusses perspectives around computability, modularity/compositionality, and intrinsic universality of networks. 
These are certainly, like the ever-growing understanding of the influences of updating modes, among the most promising tracks for future works in the field.\bigskip

\paragraph{Acknowledgement}
This work was funded mainly by our salaries as French or German State agents or 
pensioner (affiliated to Universit{\'e} Grenoble-Alpes (JD), Universit{\'e} 
d'{\'E}vry (TM and DR), Freie Universit{\"a}t Berlin (MN), and Universit{\'e} 
d'Aix-Marseille (SS)), and secondarily by the ANR-18-CE40-0002 FANs project (SS).

\bibliographystyle{alpha}
\bibliography{00_cycles}

\end{document}